\documentclass{ieeeaccess}
\usepackage{cite}
\usepackage{amsmath,amssymb,amsfonts}
\usepackage{algorithmic}

\usepackage{url}
\usepackage[colorlinks=true,linkcolor=blue,anchorcolor=black,citecolor=teal,filecolor=black,menucolor=black,runcolor=black,urlcolor=black]{hyperref}
\usepackage{amsmath}
\usepackage{tabularx}
\usepackage{xcolor}
\usepackage{soul}
\usepackage{array}
\usepackage{comment}
\usepackage{fourier} 
\usepackage{array}
\usepackage{makecell}
\sethlcolor{yellow}

\usepackage{nomencl}
\makenomenclature
\usepackage{etoolbox}

\usepackage[tight,footnotesize]{subfigure}
\usepackage{algorithm2e}
\usepackage{amsmath}
\usepackage{makecell}
\usepackage{longtable}
\usepackage{pdflscape}
\usepackage{afterpage}  
\usepackage{graphicx}
\usepackage{lipsum}
\usepackage{float}
\usepackage{multicol}
\usepackage{multirow}
\usepackage{textcomp}
\def\BibTeX{{\rm B\kern-.05em{\sc i\kern-.025em b}\kern-.08em
    T\kern-.1667em\lower.7ex\hbox{E}\kern-.125emX}}
\usepackage{changepage}
\begin{document}
\history{Date of publication xxxx 00, 0000, date of current version xxxx 00, 0000.}
\doi{10.1109/ACCESS.2017.DOI}

\title{A Physics-Informed Context-Aware Approach for Anomaly Detection in Tele-driving Operations Under False Data Injection Attacks}
\author{\uppercase{Subhadip Ghosh}\authorrefmark{}, \IEEEmembership{Senior Member, IEEE}, \uppercase{Aydin Zaboli}\authorrefmark{}, \IEEEmembership{Graduate Student Member, IEEE},
\uppercase{Junho Hong}\authorrefmark{}, \IEEEmembership{Senior Member, IEEE},
\uppercase{Jaerock Kwon}\authorrefmark{}, \IEEEmembership{Senior Member, IEEE}}
\address[]{Department of Electrical and Computer Engineering, University of Michigan -- Dearborn, MI, 48128 USA.}

\markboth
{Author \headeretal: Preparation of Papers for IEEE TRANSACTIONS and JOURNALS}
{Author \headeretal: Preparation of Papers for IEEE TRANSACTIONS and JOURNALS}

\corresp{Corresponding author: Junho Hong (e-mail: jhwr@umich.edu)}

\begin{abstract}
Tele-operated driving (ToD) systems are special types of cyber-physical systems (CPSs) where the operator remotely controls the steering, acceleration, and braking actions of the vehicle. Malicious actors may inject false data in communication channels to manipulate the tele-operator's driving commands to cause harm. Hence, protection of this communication is necessary for the safe operation of the target vehicle. However, according to the National Institute of Standards and Technology (NIST) cybersecurity framework, protection merely is not enough and the detection of an attack is necessary. Moreover, UN R155 mandates that security incidents across vehicle fleets be detected and logged. Thus, cyber-physical threats of ToD are modeled with an attack-centric approach in this paper. Then, an attack model with false data injection (FDI) on steering control commands is created from real vehicle data. The risk of this attack model is assessed for a last-mile delivery (LMD) application. Finally, a physics-informed context-aware anomaly detection system (PCADS) is proposed to detect such false injection attacks, and preliminary experimental results are presented to validate the model. 

\end{abstract}

\begin{keywords}
Tele-operated driving, Anomaly detection, Cyber-physical system, Physics-informed, Context-aware.
\end{keywords}

\titlepgskip=-15pt

\maketitle

\section{Introduction} \label{sec:introduction}
\PARstart{I}{n} recent years, autonomous driving has been one of the key areas of attention among the automotive researchers. Numerous innovations and cutting-edge technologies have emerged to bring full autonomy in road vehicles. Vehicle teleportation is one such technology that originated to provide emergency assistance to autonomous vehicles (AVs) in unusual or difficult driving scenarios~\cite{9973618, davies2018self, zaboli2023survey}. However, this technology is also being targeted for tele-operated taxis and delivery services~\cite{5GAA_ToD, 9984654, 198034, 7535475}. The National Institute of Standards and Technology (NIST) vehicle tele-operation forum and 5G blueprint project are leading the research in this area in the United States and Europe, respectively~\cite{NIST_Auto_Tele, 5gblueprint_forum}. Some start-up companies (e.g., Zoox, Ottopia, Faction, DriveU.auto) have started testing their prototypes of tele-operated vehicles for the mobility services for some specific use cases~\cite{Faction_News, driveUauto_News, Ottopia_news, Zoox_News, Vay_2024}.
\subsection{Problem Statement} \label{ProblemStatement}
In general, the driving function of a vehicle can be viewed as a combination of longitudinal control (i.e., acceleration, braking) and lateral control (i.e., steering) of a vehicle to reach from start to destination in various traffic scenarios. Tele-operated drivers can monitor, control, or provide guidance to the driving function from a remote operating station~\cite{9214882,9973618}. Typically, the perception and localization are information sent by the vehicle to the operating station via cloud and fog infrastructure using wireless or cellular networks. Similarly, control commands transmitted from the operating station are sent to the vehicle. This poses a potential exposure of perception data and control commands outside the vehicle boundary and can make the ToD vulnerable against cyberattacks. Attackers can target the ToD system with denial-of-service (DoS) attacks, FDI attacks, man-in-the-middle (MITM) attacks, and other attacks similar to the attacks detected in other CPSs~\cite{9700514, 9902043}. A malicious control of ToD may result in the vehicle crash, disruption in tele-operation service, legal consequences, and financial loss. Hence, a robust cybersecurity strategy is critical to prevent, detect, and mitigate such attacks for a safe ToD. Although cybersecurity is a common practice in information technology (IT) and many other internet of things (IoT) devices, cybersecurity for road vehicles has gained attention in recent years since a researcher in this field hacked a vehicle in 2016~\cite{Jeep_hack}. In 2020, UNECE World Forum for Harmonization of Vehicle Regulations (UNECE WP.29) has adopted UN Regulation No. 155 on Cyber Security and Cyber Security Management Systems, which requires managing cyber risks to vehicles in 54 countries from 2024~\cite{UNR155}. 
Typically, cryptography, chain of trust, firewall and access control are some of the common techniques to protect security assets in cyber domains~\cite{RAHMAN202039, 8968396, 9709835, 10090092, NSA_Strategy, AutoISAC_Sec_Dev_LifCycl, NIST_HPC_Sec}. However, with evolving threats on these methods, protection from all potential attacks cannot be guaranteed~\cite{9004435, KAUR20225766, 9311369}. Moreover, insider attacks increase the vulnerability of a system by inadequate security measures in the system design and improper implementation of cryptography algorithms which are exploited by zero-day attacks. To address this challenge, security by design needs to be followed that is the defense-in-depth (DiD) principle~\cite{NIST_DinD, OWASP_DinD}, where security strategies are applied at multiple layers. One of the critical features of DiD is the detection mechanism~\cite{nist1}. Further, UN R155 requires monitoring and reporting of security incidents for vehicle fleets for automotive applications~\cite{UNR155}. Conventional cybersecurity detection methods are primarily in the cyber domain and have limitations in addressing the security requirements of CPSs~\cite{YAACOUB2020103201, 9422483, 8617011}. To address this, recent research in other CPSs has demonstrated an extension of DiD and detection methods to physical domains~\cite{elnour2020dual, 9296745, NING2022203, wang2020anomaly, 9352761}. Currently, ToD is an emerging technology within restricted operational design domain (ODD) and prototype phase. When this technology gets deployed at large on public roads, a cyber-physical DiD strategy will be necessary to reduce risks from cyberattacks. However, to our knowledge, there is no study to show threat analysis for cyberattacks on driver's control commands transmitted from tele-operator stations to the target vehicles. Moreover, methods to detect such attacks in tele-operated vehicle's physical domain have not been explored.
\subsection{Related work}
An intrusion detection system (IDS) is one of the techniques recommended by various standards (e.g., ISO 27039, NIST, Open Web Application Security Project (OWASP)) to monitor activities in the system or network for malicious behavior. Several automotive communities and researchers are considering an automotive specific IDS as a fundamental solution for vehicle cyber incidents detection and reporting, which has the potential to be extended to intrusion detection and prevention systems (IDPSs)~\cite{NHTSA_IDS, ETAS_IDS, AutoISAC_IDS}. Automotive Open System Architecture (AUTOSAR) organization has released a specification for vehicle intrusion detection systems in 2020 that provides a standardized interface to report on-board security events for a vehicle electronic control unit (ECU) and network environment~\cite{AUTOSAR_IDS}. Basically, IDS methods in cyber domains are of three types, including signature-based, behavior-based, and anomaly-based approaches~\cite{LIAO201316, 10_1145_2808691, zaboli2023chatgpt}. Other IDS methods are inspired or combined by these basic methods. In the automotive industry, IDSs are typically software components deployed in the network, host, or as a distributed system. These types of IDS are mainly focused on messages in vehicle network protocols (e.g., CAN, automotive Ethernet)~\cite{8640808, 022_034128, 10_1145_3570954} and lack utilizing the application specific knowledge. Other than IDSs, an anomaly detection (AD) process is also used in other applications (e.g., sensor AD~\cite{rajendar2022sensor, 8684317}, vehicle traffic AD~\cite{Aboah_2021_CVPR, SARIKAN201864}, in-vehicle monitoring for AVs~\cite{app121910011}). 
In science, an anomaly is described when there is a difference between actual observation and expected outcome developed based on the original scientific idea~\cite{AnomalyDef}. In the statistics and data mining field, outliers in the dataset are considered as anomalies. 
For physical systems, detecting anomalies in AV sensors, aerial systems, and intelligent traffic systems are examples of some important applications. An AD process for IDSs was introduced in the 1980s to detect security violations by recognizing abnormal patterns in system logs~\cite{1702202}. Recent research on cyber-physical attack detection is presented in Table~\ref{tableAnomalyCPS}. According to this table, the current AD techniques for automotive IDSs primarily focus on finding anomalies based on a data-driven analysis of the network and less consideration of physical behavior. Table~\ref{tableAnomalyCPS} shows research on other CPSs found for detecting cyber-physical attacks, hybrid approaches by combining data-driven models and physics-based models. 
\begin{table*}
\centering
\caption{\centering A literature survey on AD process in CPSs. \label{tableAnomalyCPS}}
\begin{tabular}{|>{\centering\hspace{0pt}}m{0.088\linewidth}|>{\centering\hspace{0pt}}m{0.19\linewidth}|>{\centering\hspace{0pt}}m{0.144\linewidth}|>{\hspace{0pt}}m{0.513\linewidth}|} 
\hline
\textbf{Author}                                                                         & \textbf{Method}        & \textbf{Application}                                                                                          & \textbf{Contributions}                                                            \\ 
\hline
Rahul~\textit{et al.}~\cite{9064519}\textit{}~                                                 & \centering Physics guided machine learning (ML) techniques                                                                              & \textcolor[rgb]{0.2,0.2,0.2}{General CPSs}                                               & \textcolor[rgb]{0.2,0.2,0.2}{~-~}\textcolor[rgb]{0.2,0.2,0.2}{Classified the hybrid models into physics-based pre-processing, physics-based network architectures, physics-based regularization, and miscellaneous categories based on the way the model-based is brought into the hybrid architecture.~}\textcolor[rgb]{0.2,0.2,0.2}{}\par{}\textcolor[rgb]{0.2,0.2,0.2}{- Proposed five metrics for all-round performance evaluation of a hybrid CPS model.}  \\ 
\hline
Cody~\textit{et al.}~\cite{ruben2020hybrid}\par{}                                                     & \textcolor[rgb]{0.11,0.114,0.118}{Hybrid physics model-based data-driven framework}                                     & \textcolor[rgb]{0.2,0.2,0.2}{Smart grid real-time monitoring}                                                 & \textcolor[rgb]{0.2,0.2,0.2}{- Presented a hybrid framework with physics-based and data-driven ensemble CorrDet (ECD) algorithm.}\par{}\textcolor[rgb]{0.2,0.2,0.2}{- Tested the results on~}\textcolor[rgb]{0.11,0.114,0.118}{IEEE 118-bus system which shows $6.75\%$ improvement from the physic-based solution.}\textcolor[rgb]{0.2,0.2,0.2}{}                   \\ 
\hline
Faris~\textit{et al.}~\cite{9061150}\textit{}~                                                  & Statistical learning and kinematic model                                                                                & Adaptive cruise control for AVs         & - Proposed generalized extreme studentized deviate with sliding chunks (GESD-SC) approach, which is applied at each vehicle in the platoon to detect anomalies in real-time based on the vehicle's own speeding decisions.                                                \\ 
\hline
Jie~\textit{et al.}~\cite{9667119}\par{}                                                       & \textcolor[rgb]{0.2,0.2,0.2}{Spatio-temporal correlation based a long short-term memory (LSTM) method}                    & \textcolor[rgb]{0.2,0.2,0.2}{Unmanned Aerial Vehicles (UAVs)}                                                         & \textcolor[rgb]{0.133,0.133,0.133}{- Suggested an~}\textcolor[rgb]{0.2,0.2,0.2}{An spatio-temporal convolutional (STC)-LSTM algorithm which can accurately locate the anomalies of UAV flight data and provide high-precision recovery prediction values.}   \\ 
\hline
\multicolumn{1}{|>{\hspace{0pt}}m{0.088\linewidth}|}{Bin~\textit{et al.}~\cite{9743327}} & \multicolumn{1}{>{\hspace{0pt}}m{0.19\linewidth}|}{\textcolor[rgb]{0.2,0.2,0.2}{~~Physics-informed neural networks (PINNs)}} & \multicolumn{1}{>{\hspace{0pt}}m{0.144\linewidth}|}{\textcolor[rgb]{0.2,0.2,0.2}{~~~~Power systems}} & - Several paradigms of PINNs (e.g., PI loss function, PI initialization, PI design of architecture, and hybrid physics-DL models) are summarized.           \\
\hline
\end{tabular}
\end{table*}
The recent growth in ML research and its applications is largely driven by two key factors. Firstly, the digital creation and storage of extensive datasets plays a crucial role. Secondly, the accessibility of cost-effective high-performance computing devices that can process these extensive datasets acts as a vital accelerator. These datasets are often developed for particular applications, including prediction, recognition, recommendation systems, and language processing~\cite{sarker2021machine}. 
Solaas~\textit{et al.}~\cite{solaas2024systematic} performed a comprehensive literature review encompassing 203 papers concerning anomaly detection in Connected and Autonomous Vehicles. Their study highlighted LSTM, CNN, and autoencoders as the primary AI techniques and delved into the training methodologies and evaluation metrics utilized. Their evaluation revealed significant limitations: notably, only 9 out of 203 studies offered open-source availability; there was a deficiency in real-world deployment data; and there was an absence of standardized benchmarking datasets. Moreover, the research did not delve into the vulnerabilities associated with on-demand tele-operation or the use of mission-specific driving contexts for context-aware detection. Additionally, it did not investigate approaches informed by physics for the validation of vehicle behavior signatures. Mansourian~\textit{et al.}~\cite{mansourian2023deep} developed a framework for forecasting temporal events that utilizes LSTM and ConvLSTM models to detect anomalies in Controller Area Networks (CAN) through the analysis of patterns across both space and time. This approach showcased remarkable accuracy when tested on established datasets. However, the supervised approach limits flexibility against new attacks, overlooks vulnerabilities related to remote operations, and fails to incorporate validation within the context of specific missions. Additionally, the system lacks physics-based behavioral authentication and addresses only internal network security rather than comprehensive remote operation threats. A physics-informed anomaly detection framework by Guo~\textit{et al.}~\cite{guo2025physics} embedded UAV dynamics into neural detection models, demonstrating enhancements in performance, achieving increases of up to $17.77\%$ in ROC-AUC scores in countering spoofing attacks. Despite this, the approach continues to be limited to the validation of spoofing incidents and wind disturbance, failing to address the vulnerabilities associated with remote operations and the integration of operation contexts that are specific to particular routes. Moreover, while the efficiency of training is enhanced by smoothing the loss landscape, the framework did not include thorough physics-based behavioral verification and primarily targets internal UAV anomalies, neglecting the broader range of threats related to remote operations. Makridis~\textit{et al.}~\cite{makridis2023adaptive} introduced an adaptive physics-informed model that reconstructs vehicle paths by integrating constraints from vehicle dynamics with patterns of driver behavior, effectively filtering out noise from sensor data. However, their methodology focuses primarily on smoothing trajectories in an offline manner, rather than the crucial real-time anomaly detection necessary for the ToD security. Although they utilize constraints grounded in physics with effectiveness, their system lacks context recognition to determine that vehicle tasks correspond with the prescribed mission pathways. Moreover, the framework considers all anomalies to be sensor noise, neglecting to account for malicious cyber-physical threats such as FDI attacks targeting steering mechanisms. While their method shows promise for trajectory reconstruction, it necessitates precise vehicle specifications, which might not be obtainable in real-world ToD implementations. Furthermore, it lacks threat modeling and risk assessment features, which are essential for securing connected vehicle systems. Shi~\textit{et al.}~\cite{shi2021physics} developed a physics-informed deep learning (PIDL) framework with fundamental diagram learning (i.e., PIDL + FDL) for estimating traffic states and learning flow-density relationships in highway scenarios. While their methodology focuses on typical traffic reconstruction, it did not tackle the unique security issues associated with ToD. Although the framework successfully integrates physics-based models with neural networks, it fails in validating context specific to the mission and neglects considerations for harmful cyber-physical threats such as FDI attacks on steering controls. While appropriate for highway traffic analysis, it lacks both the threat assessment and real-time anomaly detection needed to protect ToD systems against adversarial manipulations. A physics-informed learning framework for autonomous screw-driving proposed by Manyar~\textit{et al.}~\cite{manyar2023physics} that characterizes rotational motion dynamics and handles position through active and passive compliance mechanisms. Although their method effectively incorporates physics-based modeling to ensure dependable assembly processes in the presence of positional uncertainties, it focuses on automating manufacturing instead of addressing ToD security issues. While the architecture includes mechanisms for identifying mechanical failures, such as cross-threading and jamming, it is deficient in functionalities for detecting sophisticated cyber-physical threats, specifically FDI attacks targeting the steering commands. Fan~\textit{et al.}~\cite{fan2024novel} constructed an advanced anomaly detection framework utilizing unsupervised Generative Adversarial Networks (GANs) in combination with LSTM networks. The proposed framework is meticulously crafted to identify adversarial threats directed at trajectory prediction algorithms. It accomplishes this through an extensive evaluation of two types of losses: the reconstruction loss, which measures the performance of the model in reproducing input data, and the discrimination loss, which assesses the model's capability to distinguish between legitimate and adversarial inputs. Nonetheless, their approach primarily concentrates on detecting malicious trajectories specific to prediction models, without fully addressing the protection of ToD operations in a holistic manner. Although the method proficiently detects adversarial trajectories by examining temporal-spatial characteristics, it is deficient in context-aware validation that would confirm maneuvers against planned mission paths. Further, the framework fails to tackle FDI attacks aimed at steering directives and lacks a physics-based validation mechanism for vehicle behavior verification. According to the provided challenges and gaps in this domain, the contributions of this research will be presented in the next section.
\subsection{Contributions}
\begin{figure}[!h]
    \centering
    \includegraphics[width=1.0\linewidth]{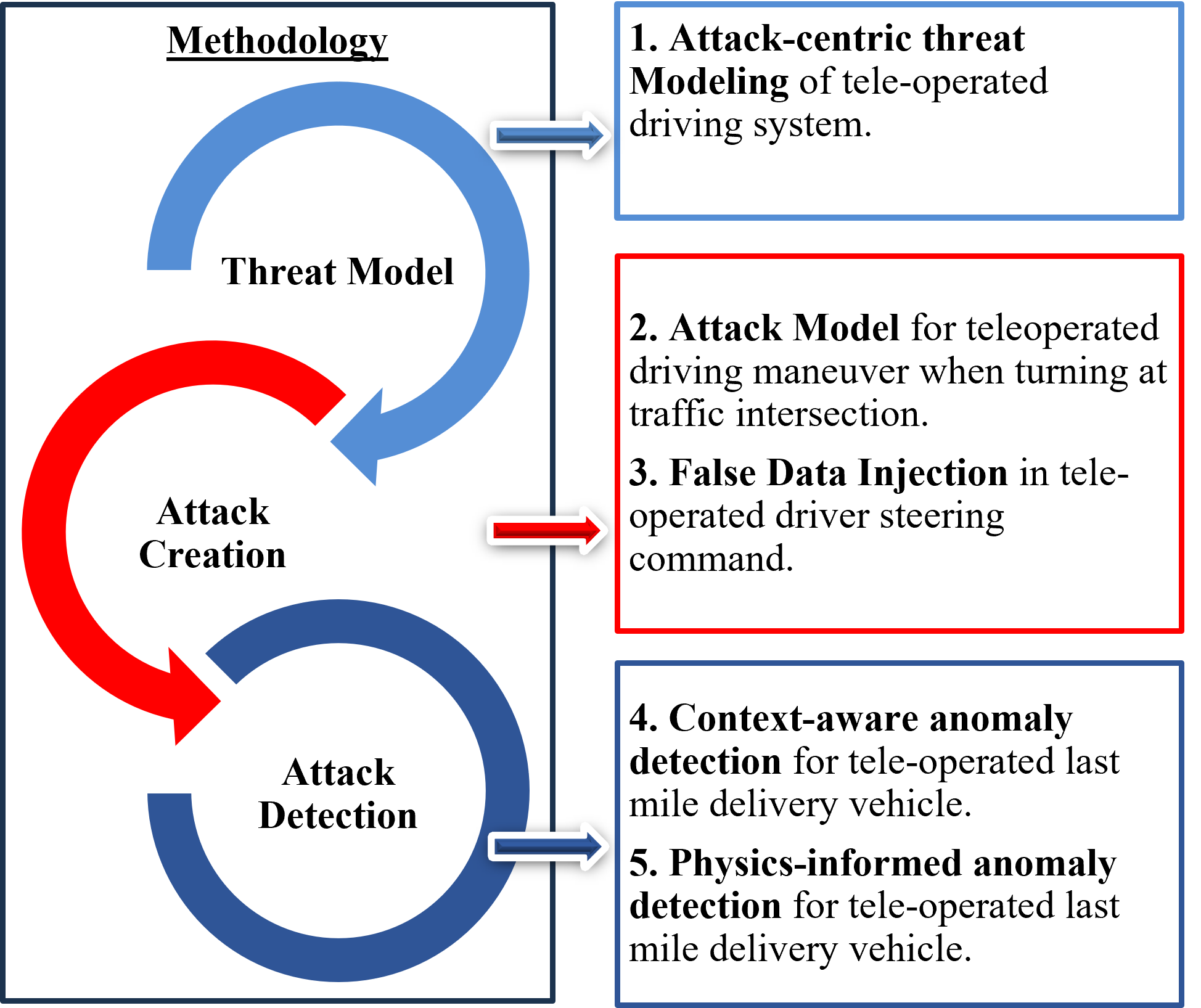}
    \caption{A workflow of the paper's contributions.}
    \label{fig:contribtuion}
\end{figure}

The primary contribution of this work is the development of a novel Physics-informed Context-Aware Anomaly Detection System (PCADS), designed to secure Tele-operation on Demand (ToD) systems against critical cyber-physical threats. The workflow culminating in these contributions is illustrated in Fig.~\ref{fig:contribtuion}.
To establish the necessity for this system, this paper first introduces a foundational threat model for ToD, an area previously unaddressed in the literature. This analysis identifies False Data Injection (FDI) on steering commands as a high-risk vulnerability. Building on this, we contribute a detailed FDI attack model, formulated and implemented by injecting noise into steering data from the D2CAV real-world driving dataset during turning maneuvers~\cite{toghi2020maneuverbased}. The core contribution is the PCADS framework itself, which pioneers a dual-pronged detection strategy by integrating two innovative concepts: a context-aware module that leverages the vehicle’s mission-specific Driving Contexts (DCs) and a physics-informed module that learns the vehicle's physical response signatures during maneuvers. The principal contributions are therefore:
\begin{itemize}
    \item \textbf{A foundational cyber-physical threat analysis and risk assessment for ToD systems,} identifying previously uncatalogued vulnerabilities.
    \item \textbf{A novel FDI attack model targeting steering control,} completed with its mathematical formulation and validation on real-world driving data.
    \item \textbf{A context-aware anomaly detection method} that uniquely utilizes mission-specific driving contexts to validate vehicle maneuvers against intended routes.
    \item \textbf{A physics-informed anomaly detection method} that learns and verifies the physical signatures of vehicle behavior, providing a robust, model-based layer of security.
\end{itemize}
\subsection{Assumptions and Scope}
\begin{itemize}
    \item This work is focused on a specific use case of ToD which is last-mile delivery (LMD).
    \item For DCs, a solution of the vehicle routing plan (VRP) to find optimal routes for the fleet of vehicles is not in the scope of this paper and it is assumed the VRP is accurate and robust to address real-time traffic density, road conditions, weather and vehicle maintenance schedule.
    \item A dynamic alert generation is out of scope in this research which it is simulated as a binary flag.
    \item For the physical parameter learning, left turn, right turn and U-turn maneuvers are considered.
    \item Experimental results are based on the dataset mentioned in experiment section.
    \item The proposed methodology assumes the vehicle configuration and physical parameter values for left turn, right turn and U-turn maneuvers of the target vehicle that are known to AD system. 
\end{itemize}

\subsection{Paper Structure}
The rest of this paper is organized as follows: Section~\ref{sec:TARA} provides cyber-physical threat models for ToD and attack models for injecting noise on steering control command for left turn, right turn, and U turn actions at a traffic intersection. Section~\ref{sec:PCADS} describes the proposed AD method, PCADS, and the corresponding mathematical modeling. The experimental setup and results are discussed in Section~\ref{sec:experiment}. Finally, the paper is concluded in Section~\ref{sec:conclusion}, and the supplementary material is given in Section~\ref{appendix-1}.

\section{A ToD Threat Analysis}\label{sec:TARA}
Tele-operation on Demand (ToD), or tele-operation, provides remote driving or assistance to piloted and autonomous vehicles (AVs) and is a critical component of the AV ecosystem, enabling services like tele-operated taxis and deliveries~\cite{Telops_Press, NIST_Telop_Conf, 5G_Remote_Station}. This technology allows a remote operator to passively monitor and, if necessary, take full control of semi-autonomous and autonomous vehicles. The automotive industry categorizes ToD into three main types of control. As described in Table~\ref{Tab:ToD_Func}, in direct control, the remote operator manages most driving functions, including planning and decision-making. With indirect control, the operator guides the vehicle by providing or selecting trajectories. Shared control involves a division of decision-making and vehicle control between the automated driving system and the remote operator~\cite{Teleops_Survey, HMI_Robot_Taxonomy}.
\begin{table*}  
\centering
\caption{\centering Functions and data flows for various ToD events.}\label{Tab:ToD_Func}
\resizebox{\textwidth}{!}{%
\begin{tabular}{|>{\centering\arraybackslash}m{0.12\linewidth}|>{\centering\arraybackslash}m{0.12\linewidth}|>{\centering\arraybackslash}m{0.15\linewidth}|>{\centering\arraybackslash}m{0.18\linewidth}|>{\centering\arraybackslash}m{0.06\linewidth}|>{\centering\arraybackslash}m{0.105\linewidth}|>{\centering\arraybackslash}m{0.105\linewidth}|>{\centering\arraybackslash}m{0.09\linewidth}|>{\centering\arraybackslash}m{0.09\linewidth}|>{\centering\arraybackslash}m{0.13\linewidth}|>{\centering\arraybackslash}m{0.09\linewidth}|} 
\hline
\multicolumn{2}{|c|}{} & \multicolumn{2}{c|}{\textbf{Data Flow}} & \multicolumn{7}{c|}{\textbf{Function}} \\ 
\hline
\multicolumn{2}{|c|}{} & \textbf{Vehicle to Operating Station} & \textbf{Operating Station to Vehicle} & \textbf{Sensing} & \textbf{Perception} & \textbf{Localization} & \textbf{Planning} & \textbf{Decision} & \textbf{Control} & \textbf{Actuation} \\ 
\hline
\multicolumn{2}{|c|}{\textbf{Direct Control}} & Sensor data & \makecell{Control command via \\steering, brake, acceleration\\ pedal operation by \\remote operator.} & Vehicle & Operator station & Operator station & Operator station & Operator station & \makecell{Hi level: \\Operator station\\ Lo level:\\ Operator station} & Vehicle \\ 
\hline
\multicolumn{2}{|c|}{\makecell{\textbf{Shared Control}}} & \makecell{Object list or a \\representation of the \\free space.} & \makecell{Desired control command \\via steering, brake,\\ acceleration pedal \\operation by remote \\operator.} & \makecell{Vehicle} & \makecell{Vehicle} & Vehicle & Operator station & Vehicle and Operator station & \begin{tabular}[c]{@{}c@{}}\makecell{Hi level: \\ Vehicle Operator \\station\\ Lo level: Vehicle}\end{tabular} & Vehicle \\ 
\hline
\multirow{4}{*}{\textbf{Indirect Control}} & \textbf{Trajectory Guidance} & Sensor data & Control command as trajectory & Vehicle & Vehicle/Operator station & Vehicle/Operator station & Operator station & Operator station & \begin{tabular}[c]{@{}c@{}}\makecell{Hi level: Operator \\ station\\ Lo level: Vehicle}\end{tabular} & Vehicle \\ 
\cline{2-11}
 & \textbf{Waypoint Guidance} & Sensor data & Discrete waypoints & Vehicle & Vehicle/Operator station & Vehicle/Operator station & Operator station & Operator station & Vehicle & Vehicle \\ 
\cline{2-11}
 & \textbf{Interactive Path Planning} & Object list and a grid map & Optimized path & Vehicle & Vehicle & Vehicle & Vehicle and Operator station & Operator station & Vehicle & Vehicle \\ 
\cline{2-11}
 & \textbf{Perception Modification} & Object list and a grid map & Bounding box & Vehicle & Vehicle and Operator station & Vehicle & Vehicle & Vehicle & Vehicle & Vehicle \\ 
\hline
\end{tabular}
}
\end{table*}
\subsection{Threat Modeling of Tele-operated driving}
\begin{figure}[!h]
\centerline{\includegraphics[width=1.0\columnwidth]{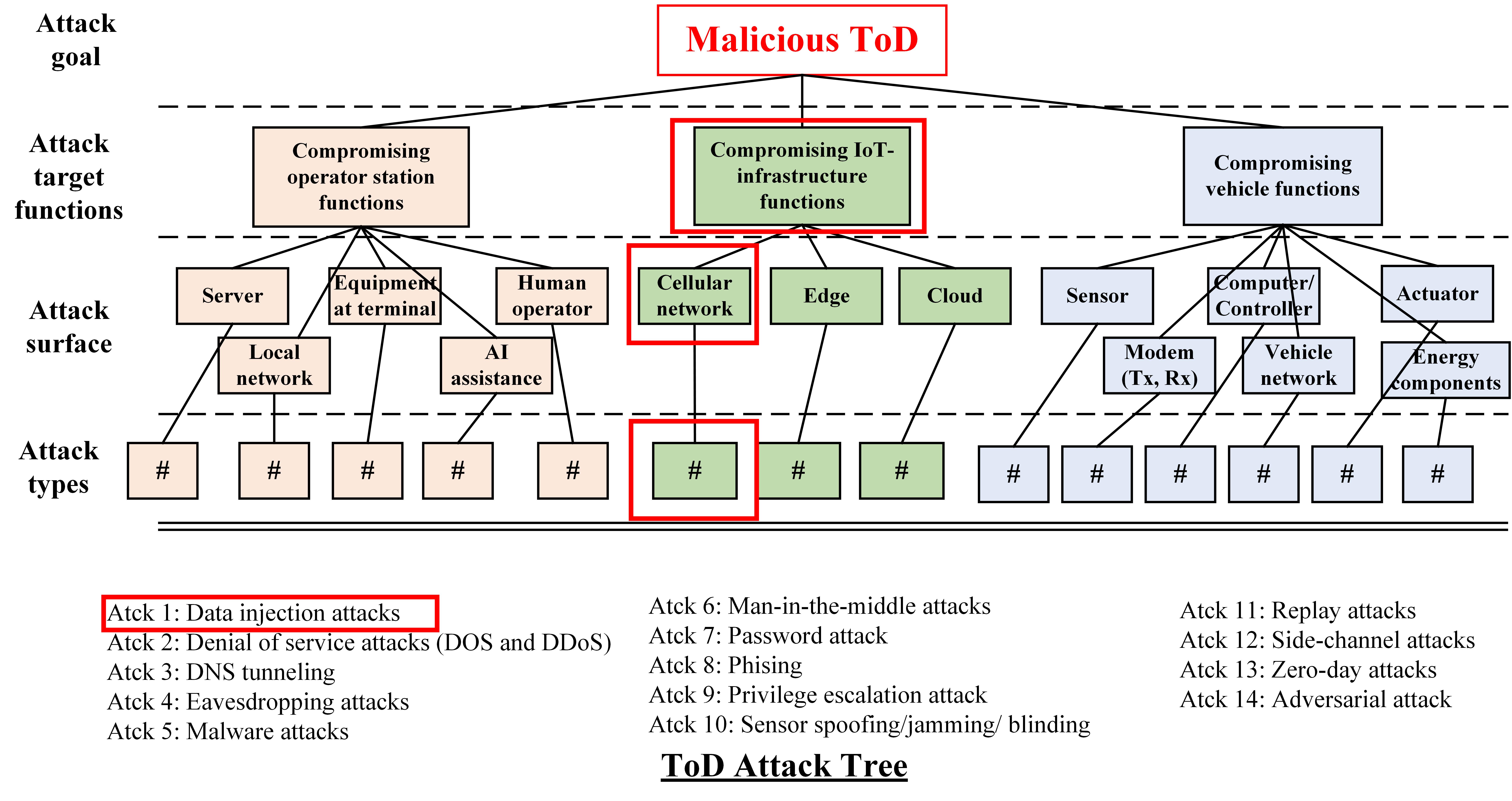}}
\caption{An attack tree for a ToD event.}
\label{TelopsAtkTree}
\end{figure}

\begin{figure}[!h]
\centerline{\includegraphics[width=1.0\columnwidth]{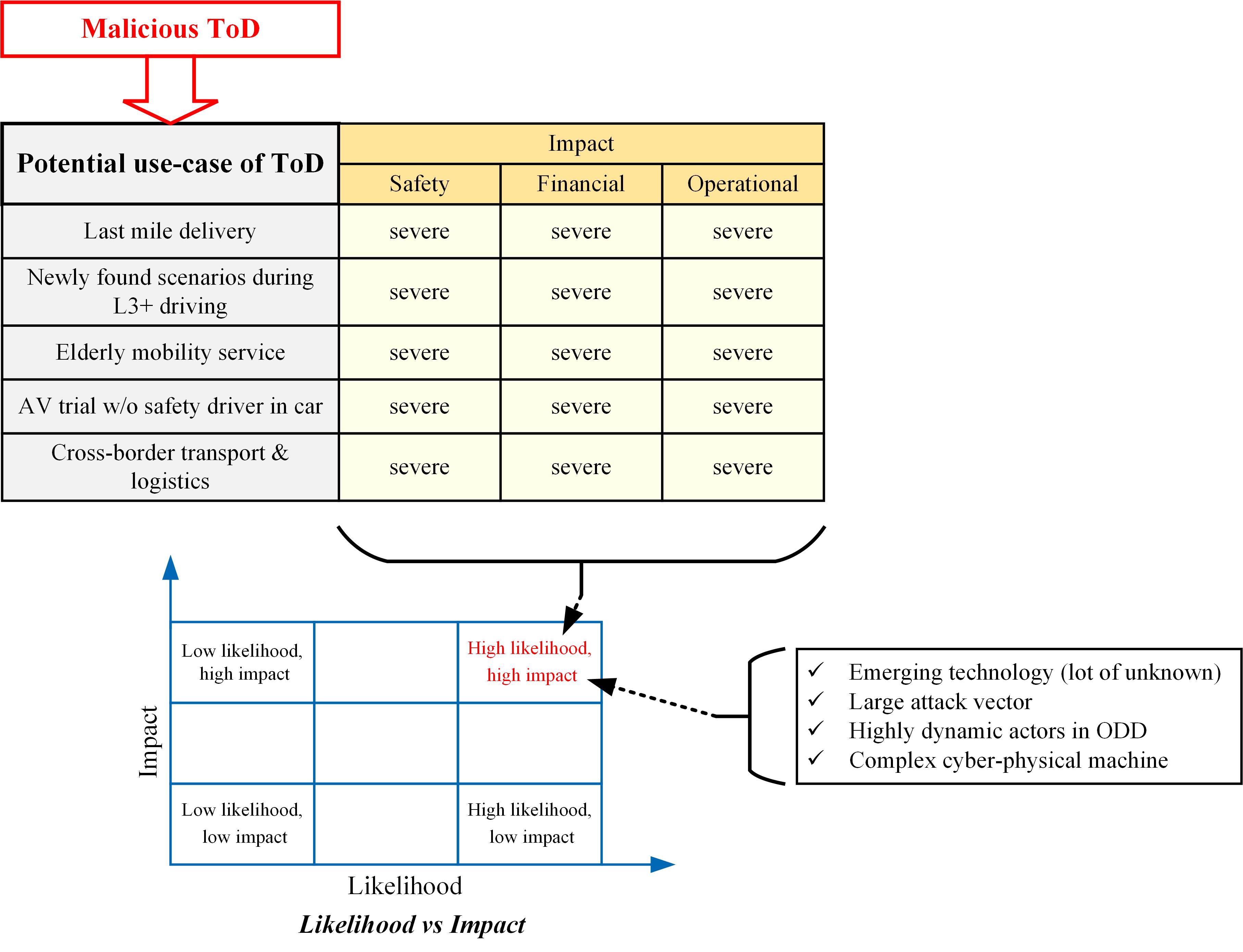}}
\caption{A tele-operated vehicle attack likelihood vs impact.}
\label{LikelihoodImpact}
\end{figure}

A threat modeling, also referred to as threat analysis and risk assessment (TARA) model, is generally viewed as the starting point for designing a cyber-secure system~\cite{ghosh2023integrated}. Given the extensive and dispersed attack surface of vehicle tele-operation, an attack-tree based method is employed for the threat analysis. In this study, TARA of the ToD system is carried out with the following steps. The first step involves identifying the components of a ToD system. According to Table~\ref{Tab:ToD_Func}, ToD functions can be distributed in three categories of components (e.g., operator station, IoT infrastructure, and vehicle). An operator station must include human operator, operator terminal, server, and local communication network. It might also have artificial intelligence (AI) assistance to the terminal. An IoT infrastructure can be divided into three primary sub-components including cellular network, cloud, and edge. From the ToD perspective, the vehicle needs to have sensing devices for perception, localization, inertia, and vehicle diagnostics. A vehicle also requires an in-vehicle communication network to communicate between multiple ECUs and a modem to communicate using cellular channels. For vehicle motion, it needs the drive-train, controller, and actuators. In the second step, all of these components are organized in a tree format as shown in Fig.~\ref{TelopsAtkTree}. 
In the third step, the attack tree is created with potential attacks on the ToD system. The attack tree in Fig.~\ref{TelopsAtkTree} illustrates the variety of attacks that can be aimed at different components from the vehicle to the operator station, potentially leading to malicious ToD. These attacks are determined based on the literature review for other CPSs and MITRE ATT\&CK® matrices~\cite{MITRE}. 
In the $4^{th}$ step, most promising applications of ToD are analyzed for an impact on safety, finance, and legality due to malicious ToD, as illustrated in Fig.~\ref{LikelihoodImpact}. 
This analysis is carried out with a subjective approach considering that malicious ToD can disrupt lateral and longitudinal motions, and suspension control of the tele-operated vehicle. Such incidents can interrupt ToD service and even jeopardize the safety of passengers and other road users. In~\cite{asi5040082, anderson2018automated, MURIEL2022103915, penmetsa2021effects}, researchers have discussed various consequences of disrupting cross-border transportation, LMD, and mobility services. According to these papers, disruption of these applications has major adverse effects on road safety and the regional economy. As these are the potential applications for ToD, it can be argued that a malicious ToD event can inflict serious harm on these applications. Finally, in the $5^{th}$ step of the TARA, a risk is assessed based on the likelihood of attacks causing a malicious ToD and the impact on the ToD application. At present, ToD for public roads is still a developing technology. However, when ToD is implemented on public roads, attack surface and the number of impacted users will expand. As a result, the risk will also escalate, which is demonstrated as a high risk in Fig.~\ref{LikelihoodImpact}.

\subsection{Attack Model for Last-Mile Delivery}
\begin{figure}[!h]
\centerline{\includegraphics[width=1.0\columnwidth]{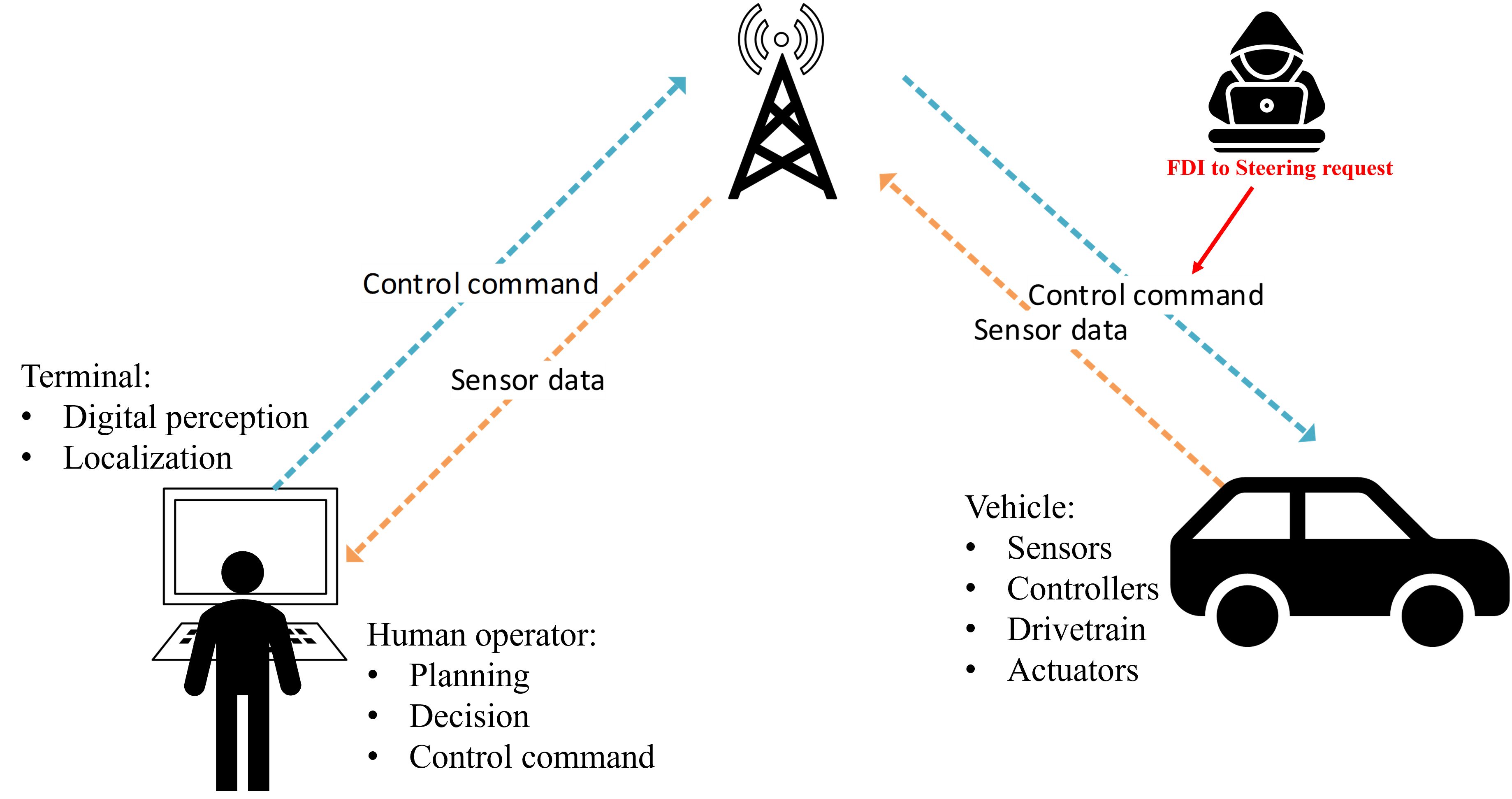}}
\caption{An FDI attack on communication between remote operators and the vehicle.}
\label{AtkRemoteVehComm}
\end{figure}

\begin{figure}[!h]
\centerline{\includegraphics[width=0.75\columnwidth]{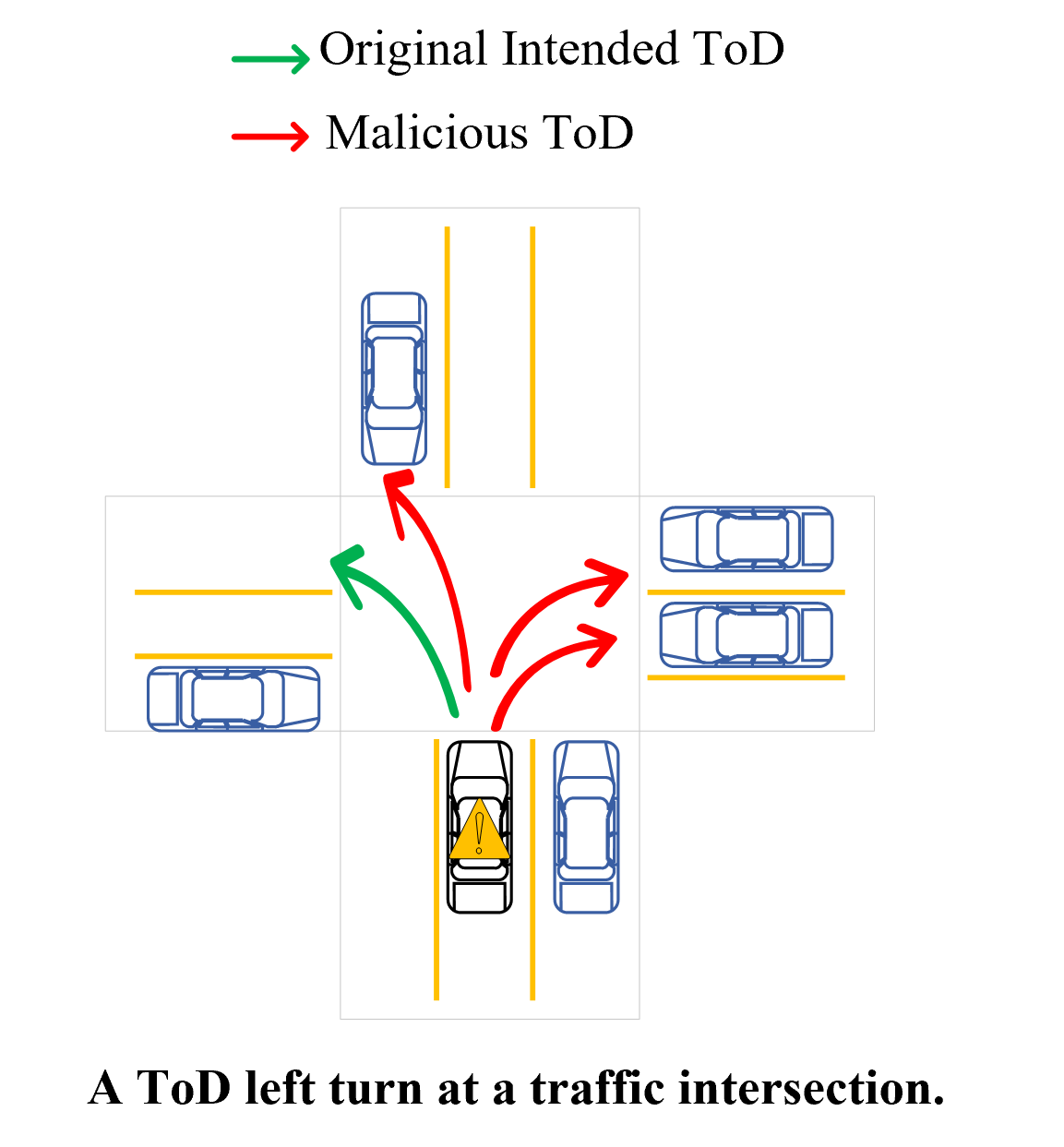}}
\caption{A traffic light intersection attack scenario.}
\label{Scenario}
\end{figure}

\begin{figure}[!h]
\centerline{\includegraphics[width=1.0\columnwidth]{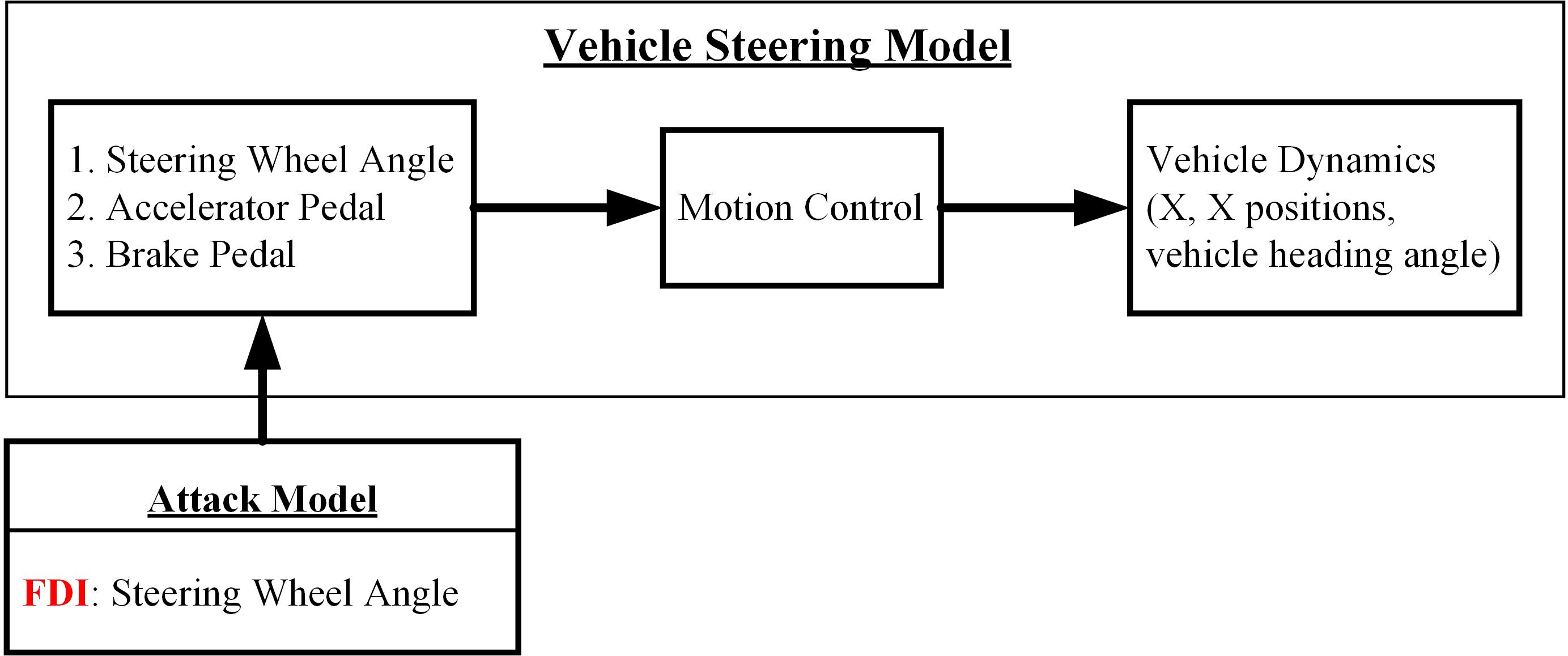}}
\caption{A system model for the vehicle's steering angle under attacks.}
\label{SystemMdl}
\end{figure}

\begin{figure}[!h]
\centerline{\includegraphics[width=1.0\columnwidth]{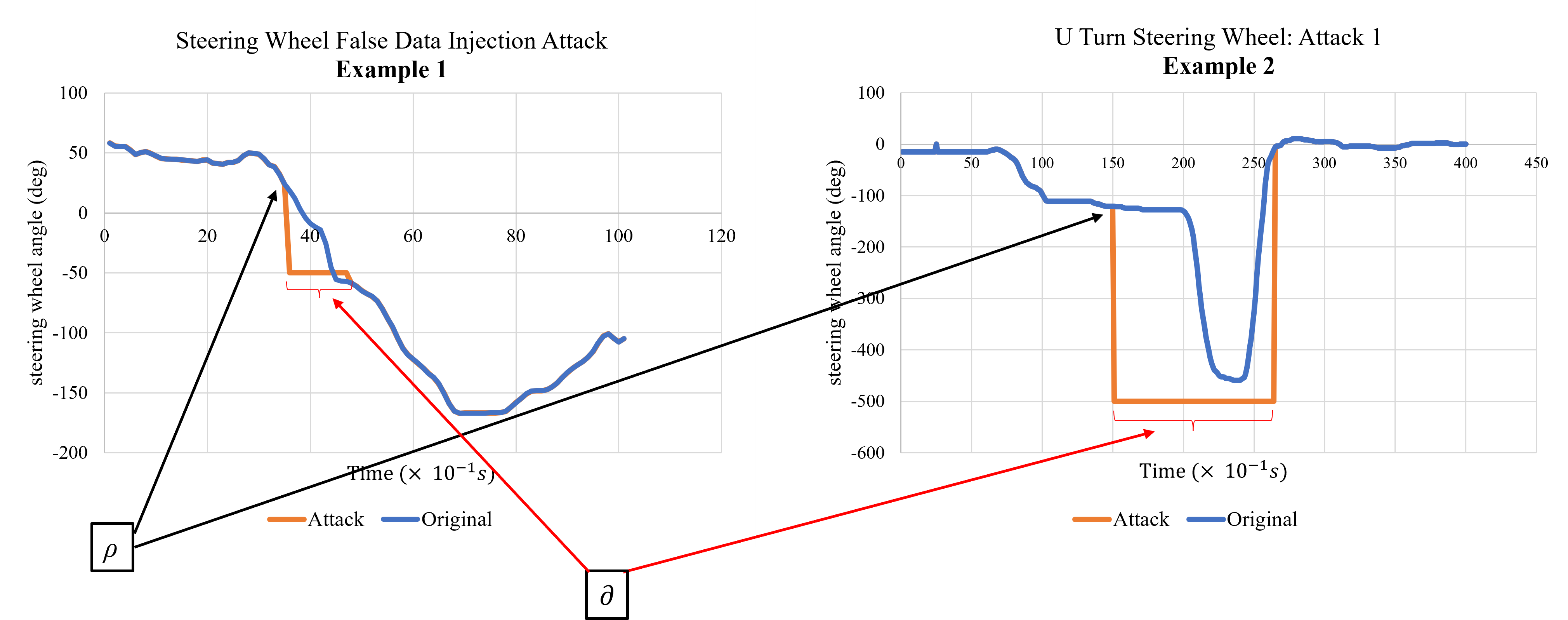}}
\caption{FDI attacks on steering wheel angle commands demonstrating two distinct attack scenarios with varying injection points and durations.}
\label{AttackMath}
\end{figure}

According to the attack tree analysis, an attacker can cause a malicious ToD event by compromising the IoT infrastructure. In this section, firstly, an attack model is developed for one of the potential use cases of the ToD event on public roads, known as ``last-mile delivery (LMD)''. Secondly, an attack formulation is implemented for an FDI attack on the steering wheel angle command from the tele-operator. The LMD represents the concluding stage in a business-to-customer (B2C) delivery process where the package is transported to the recipient, either directly to their home or to a designated pickup location~\cite{tiwapat2018last}.
Fig.~\ref{AtkRemoteVehComm} illustrates the performance of the tele-operated LMD vehicle.
This illustration showcases the essential framework of the remote vehicle operation system, which serves as the primary motivation for this study into tele-operated vehicles. The inclusion of this diagram is essential to understanding the attack surface which is investigated throughout this work. The system functions within a two-way communication framework, wherein the vehicle's onboard sensors consistently gather environmental data and transmit this information to a remote operator station through cellular or wireless networks. At the terminal, the human operator leverages advanced digital perception systems and localization technologies to evaluate the driving environment. Utilizing this remote visual interface, the operator conceives navigational strategies and develops control instructions, which are conveyed to the vehicle via the identical communication framework. Upon receiving these directives, the vehicle's integrated systems which include controllers, drivetrain components, and actuators, carry out the designated movements. Critically, this architectural design reveals a potential security weakness, as indicated by the red arrow. FDI attacks can target steering control commands during transmission. This security risk grows increasingly relevant in LMD scenarios, where AVs regularly maneuver through complex urban landscapes that demand accurate steering modifications, particularly during turning maneuvers at intersections as they follow specified delivery pathways~\cite{goodall2020non}.
However, an attack on the steering command can execute an undesired driving action and trajectory, causing an accident that is depicted in Fig.~\ref{Scenario}. 
According to this diagram, the normal trajectories for vehicle \textit{A} are shown with green arrows, but the vehicle follows the path shown in red arrows under a cyberattack. This malicious behavior can be a potential cause of frontal or angled collisions with other stationary and moving road users. Based on US NSC data 2020, angled collisions and head-on collisions are the top two reasons for deaths and fatal crashes in the U.S.~\cite{NSCCrashType}. The analysis of damage patterns and severity of impact for passenger cars presented by Kurebwa \textit{et. al} shows that the probability of damage and severity is significantly higher at the front and front corner zones as compared to other points of impact on a vehicle~\cite{3927935}. Hence, an FDI attack on the steering command from the tele-operator is selected for the case study of the attack model. An attack model designed for this study is presented in Fig.~\ref{SystemMdl}. 
As depicted, driver inputs for steering wheel angle, accelerator pedal, and brake pedal determine the motion control logic of a tele-operated vehicle. Furthermore, motion control signals determine the vehicle heading angle and vehicle dynamics. Therefore, it can be derived that an FDI attack on driver input for steering wheel angle will impact the vehicle heading angle and dynamics. In order to create this attack, an attack formula is developed for the FDI on steering wheel angle. For this purpose, ISO/SAE $21434$ is reviewed for recommended core factors to assess the attack feasibility. 
Empirical data illustrating the execution and effects of FDI attacks on steering wheel angle directives are depicted in Fig.~\ref{AttackMath}.
Incorporating this figure is crucial for illustrating the dynamics of the attack and for clarifying the practical implications embedded in this theoretical framework. The figure illustrates two separate attack scenarios, each demonstrating a unique strategy of exploitation. In Example 1, a moderate steering maneuver is observed where the vehicle's steering angle transitions from approximately 50 degrees to -170 degrees. The attack signal (depicted in orange) introduces a sharp rectangular pulse deviation at the critical moment when the steering angle begins its reduction, specifically around the 30--40 time unit mark. The timing strategy illustrates the method by which adversaries can take advantage of transitional steering phases to achieve maximum disruption while minimizing the duration of the injection. Example 2 demonstrates a more intense situation featuring a U-turn maneuver, characterized by steering angles that vary between 0 and -450 degrees. In this scenario, the attack is characterized by the injection of a continuous rectangular pulse within the 150 to 250 time unit interval, resulting in an extended simulated steering command of -500 degrees. This instance showcases that extended attack durations may be utilized in complex maneuvers to possibly trigger significant deviations in the trajectory. These attacks are grounded in a mathematical framework that is effectively represented through Eq.~(\ref{eq:Attackformulation}), where $\hat{\alpha}$ represents the falsified steering command, $\alpha$ denotes the legitimate command, $\epsilon$ indicates the injected fault magnitude, and $\Bar{\omega}$ defines the window of opportunity (\textit{WoO}) as a function of the injection point ($\rho$) and duration.
\begin{equation} \label{eq:Attackformulation}
    \hat{\alpha} = f(\alpha, \epsilon, \Bar{\omega})
\end{equation}
The graphical visualizations reveal that an effective attack depends on the precise coordination of the injection timing with the dynamics of the vehicle's steering. The varying approaches, i.e., short and precisely timed as seen in Example 1 as opposed to the extended approach in Example 2, highlight the wide range of attack methodologies that adversaries could utilize in targeting tele-operated vehicle systems.
\section{Proposed Physics-Informed Context-Aware Anomaly Detection System}\label{sec:PCADS}
In this section, a Physics- and Context-Aware Anomaly Detection System (PCADS) is proposed to detect anomalies during left, right, and U-turns in Last Mile Delivery (LMD) vehicles, focusing on False Data Injection (FDI) attacks against the steering wheel angle. The method, outlined in Fig.~\ref{ToD_AD_Framework}, requires vehicle-specific Driving Contexts (DCs) and time-series patterns of physical parameters.
\begin{figure*}[!h]
\centerline{\includegraphics[width=2.0\columnwidth]{figures/Fig_8_PCADS_framework.jpg}}
\caption{An LMD ToD anomaly detection framework.}
\label{ToD_AD_Framework}
\end{figure*}
As depicted, the method has two stages:
\begin{itemize}
    \item A context-aware anomaly detection (PCADS-CA).
    \item A physics-informed anomaly detection (PCADS-PI).
\end{itemize} 
The PCADS-CA stage compares the intended maneuver, inferred from DCs, with the actual driving command from the teleoperator. The PCADS-PI stage monitors the vehicle's physical parameter patterns during a turn and compares them against learned patterns to detect deviations. The following sections detail the PCADS-CA and PCADS-PI models.
\subsection{Stage 1: PCADS-based Context-Aware Anomaly Detection Framework}\label{subsec:PCADS-CA}
The context-aware anomaly detection module suggests that the environmental and situational conditions around a vehicle can effectively anticipate expected driving actions. This framework establishes a predictive chain where driving contexts (DCs) inform the intended maneuver (IM), which subsequently corresponds to the actual control inputs ($D_I$) from the tele-operated driver. This hierarchical framework constitutes the foundational structure of the anomaly detection strategy. DC is mathematically expressed as Eq.~(\ref{eq:D_C}), a function of multiple environmental and operational parameters:
\begin{equation}\label{eq:D_C}
    D_C = f_1(m, \gamma, t, \omega, \tau, l, \varepsilon)
\end{equation}
Here, the DC incorporates the mission parameters ($m$), current road conditions ($\gamma$), traffic congestion levels ($t$), weather conditions ($\omega$), temporal information ($\tau$), geographical location ($l$), and various dynamic factors ($\varepsilon$) that may influence driving decisions. Furthermore, the system takes into account the predetermined route ($R$) as well as the intersection points ($f$) along the trajectory. The IM prediction follows from the DC through Eq.~(\ref{eq:D_M}):
\begin{equation}\label{eq:D_M}
    D_M \in \{st, lt, rt, ut\} = f_2(D_C)
\end{equation}
This function maps the contextual information to specific maneuver types: continuing straight ($st$), executing a left turn ($lt$), performing a right turn ($rt$), or making a U-turn ($ut$). Subsequently, these IMs translate into specific vehicle control commands that are illustrated in Eq.~(\ref{eq:D_I}):
\begin{equation}\label{eq:D_I}
    D_I \in \{Cmd_{str}, Cmd_{accl}, Cmd_{Brk}\} = f_3(D_M)
\end{equation}
These commands encompass steering inputs ($Cmd_{str}$), acceleration commands ($Cmd_{accl}$), and braking instructions ($Cmd_{Brk}$), with vehicle health status ($\hat{H}$) and context-based anomaly flags ($A'_C$) being monitored throughout. This detection framework implements three distinct verification mechanisms to identify potential security breaches or system malfunctions as follows:
\subsubsection{Incorrect Maneuver Detection at Intersections}
The initial detection mechanism continuously observes the vehicular behavior at designated intersection locations. When the vehicle ($V$) operates under an active mission ($m$), the system initializes tracking for the vehicle's lateral position (${D}_{I_{Lat}}$), longitudinal position (${D}_{I_{Long}}$), and executed maneuver (${D}_{I_{Mnvr}}$). As the vehicle navigates the planned route, the system locates all intersection features and establishes a reference database ($E$) containing expected lateral positions, expected longitudinal positions, and anticipated maneuvers. Throughout the operational phase, the system persistently evaluates the actual vehicle position and maneuver relative to the anticipated parameters. In the event that an inconsistency is identified, particularly when the maneuver executed at the current location deviates from the expected maneuver at the corresponding position, an anomaly flag ($A'_C$) is activated and documented within the database.

\subsubsection{Temporal Window Validation}
The second verification layer implements temporal constraints on the anomaly detection process. This layer enhances spatial validation by integrating time-dependent parameters ($\tau$). The system assesses whether maneuvers are executed within acceptable time frames by correlating the actual timing of maneuvers with predefined temporal limits. In instances where a vehicle executes a maneuver at the correct spatial location but beyond the expected temporal window, this temporal deviation activates an anomaly alert. This method efficiently detects attacks that might exploit timing manipulations to trigger hazardous conditions while preserving spatial integrity.

\subsubsection{Dynamic Alert Filtering and False Positive Reduction}
The third component mitigates the issue of false positives through the implementation of intelligent filtering. This mechanism considers dynamic environmental factors ($\varepsilon$) and vehicle health status ($\hat{H}$) before confirming an anomaly. The system conducts standard positional and temporal validations, further ensuring the absence of dynamic factors and confirming the vehicle's health status as indicating normal operation. The system confirms the anomaly flag exclusively when two criteria are met: there is no dynamic interference ($\varepsilon = \text{NULL}$), and the vehicle's health status is verified as satisfactory ($\hat{H} = \text{OK}$). This comprehensive verification approach effectively minimizes the occurrence of false positives, yet continues to preserve the sensitivity required to detect authentic security threats.

This extensive context-aware methodology guarantees effective anomaly detection by utilizing environmental insights, temporal limitations, and advanced filtering to differentiate between authentic operational deviations and possible security breaches within tele-operated vehicle systems.
\subsection{Stage 2: PCADS-Physics-informed anomaly Detection} \label{sec:PCADS-PI}
\begin{figure*}[!h]
\centerline{\includegraphics[width=2.0\columnwidth]{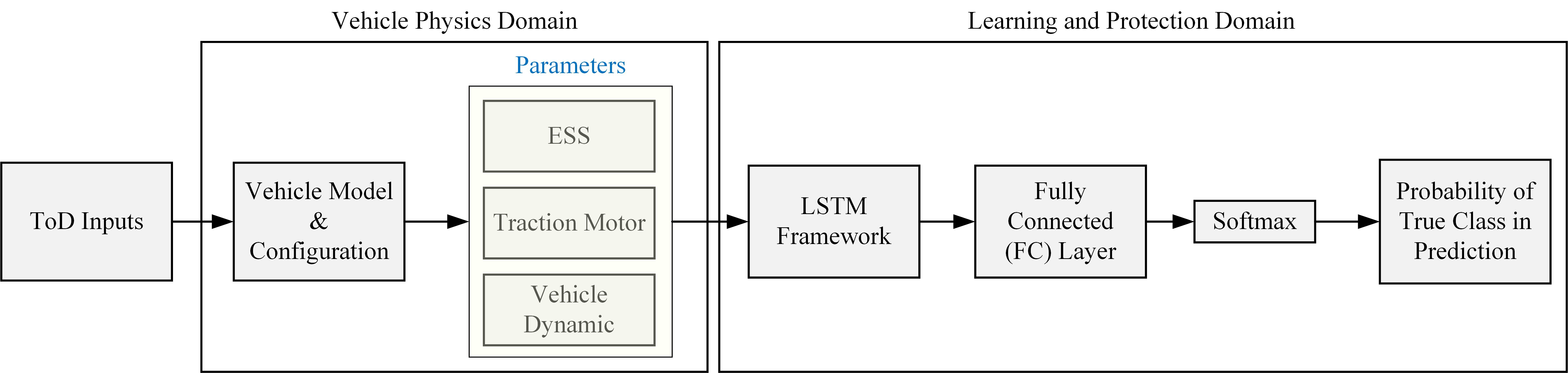}}
\caption{A vehicle physics informed anomaly detection framework.}
\label{VehPhys_AD}
\end{figure*}

The PCADS-PI method leverages the vehicle's physical elements (e.g., power transfer, vehicle dynamics) to validate cyber-element commands (e.g., steering, acceleration). As shown in Fig.~\ref{VehPhys_AD}, the detection mechanism is divided into a vehicle physics domain and a learning/prediction domain. 
ToD inputs are fed into the vehicle physics model, whose output is then passed to an ML algorithm that learns the correlation between physical responses and specific maneuvers to predict deviations.

\subsubsection{Proposed Mathematical Modeling}
The PCADS-PI framework is modeled as follows.

\begin{paragraph} {Inputs}
Time-series ToD inputs include acceleration-pedal-position (APP), steering-wheel-angle (SW), and brake-pedal-status (BP) over a time window of size N.
\begin{equation} \label{APP}
    \text{APP} \rightarrow \{APP_{t-N},..., APP_{t}\}
\end{equation}
\begin{equation} \label{SW}
    \text{SW} \rightarrow \{SW_{t-N},..., SW_{t}\}
\end{equation}
\begin{equation} \label{BP}
    \text{BP} \rightarrow\{BP_{t-N},..., BP_{t}\}
\end{equation}
\end{paragraph}

\begin{paragraph} {Vehicle Model \& Configuration}
The vehicle model translates inputs into motion based on its configuration, including drivetrain ($D$), steering system, and tires.
\begin{itemize}
    \item \textbf{Drivetrain Dynamics:} $T_{output} = f(APP_{t}, D)$
    \item \textbf{Steering Dynamics:} $\theta_{steer} = g({SW}_{t})$
    \item \textbf{Braking Dynamics:} $a_{decelaration} = h({BP}_{t})$
\end{itemize}
The overall motion is determined by integrating these dynamics into the vehicle's equations of motion:
\begin{equation}
\begin{split}
    \frac{d(\text{Vehicle State})}{dt} = \Psi (T_{output}, \theta_{steer}, a_{deceleration}, \\  \text{Vehicle Configuration})
    \end{split}
\end{equation}
\end{paragraph}

\begin{paragraph} {Model Parameters}
The model integrates control inputs with physical responses from the Energy Storage System (ESS), Motors (M), and Vehicle Dynamics (VD).
\begin{itemize}
    \item \textbf{ESS Dynamics:} Battery power $P_{battery}$ is the sum of motor power consumption.
    \item \textbf{Motor Dynamics:} Torque and speed are functions of APP and vehicle state.
    \item \textbf{Vehicle Dynamics:} Parameters like Wheel Angle (WA), Roll (R), Pitch (PT), and Yaw are derived from steering, braking, and acceleration forces.
\end{itemize}
\end{paragraph}
\subsubsection{Machine Learning Framework}
\begin{figure}[!h]
\centering
\includegraphics[width=1.0\linewidth]{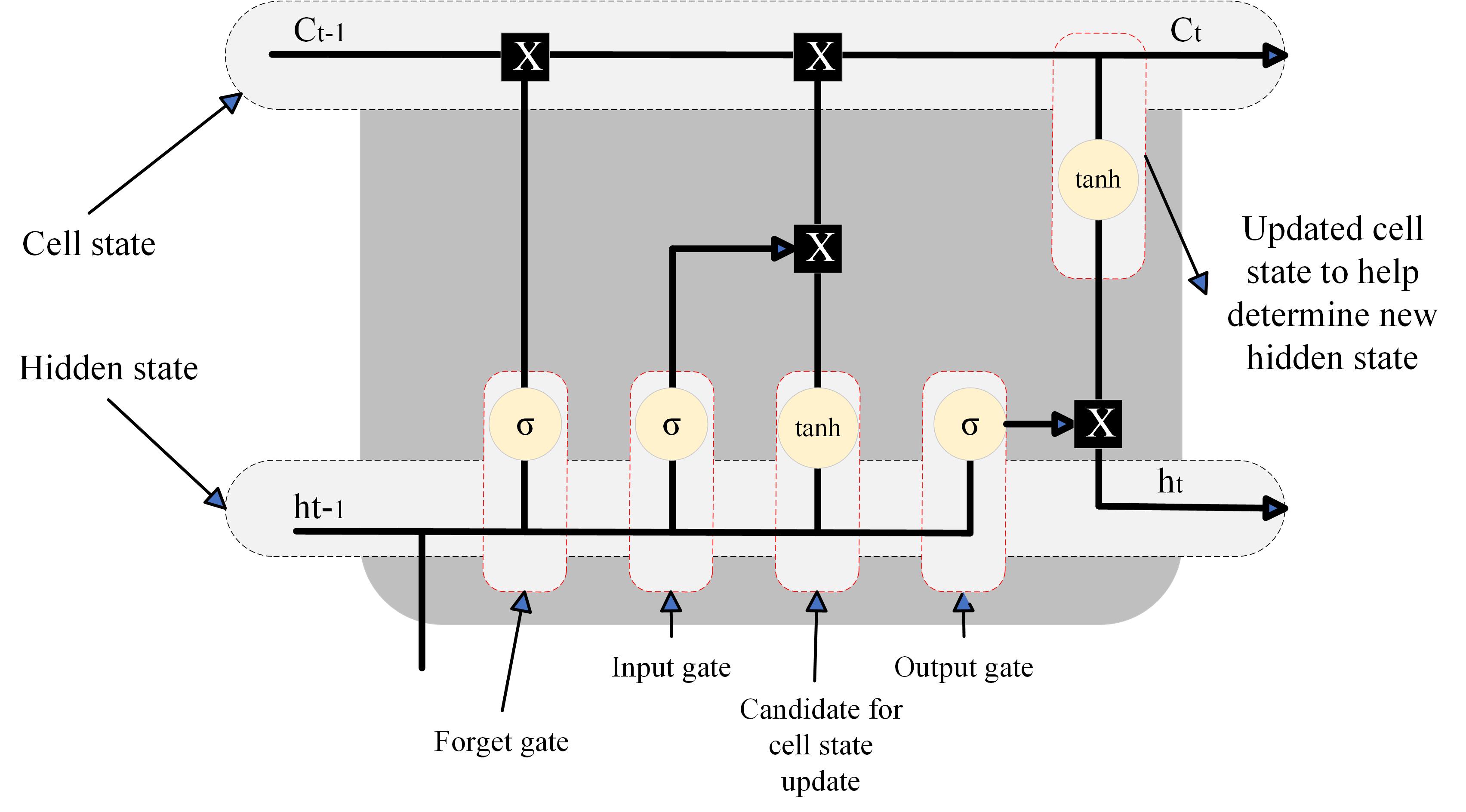}
\caption{A standard LSTM cell showing forget, input, and output gates with $\sigma$/tanh activations and state propagation paths~\cite{zaboli2024comprehensive}.}
\label{fig:LSTM}
\end{figure}

Various ML methods were reviewed, including Naive Bayes, Decision Tree, SVM, and KNN. The Long Short-Term Memory (LSTM) algorithm, a type of RNN, was selected as the base model due to its efficacy in learning from complex sequential data~\cite{li2023scenarios, chu2023review}. An LSTM unit (Fig.~\ref{fig:LSTM}) uses a cell state and gates (input, forget, output) to regulate information flow, enabling it to capture long-term dependencies in time-series data.
For this research, the LSTM model is applied to data from a full-electric vehicle model, using parameters from the ESS and traction motors.
\paragraph{A Vehicle Model Integration into the LSTM Framework}
The architecture for processing temporal sequences employs an LSTM, which at each timestep, $t$, processes a feature vector as its input, defined as $\mathbf{x}_t = [E_t, M_t, VD_t]$. These components represent the ESS parameters ($E_t$), motor characteristics ($M_t$), and vehicle dynamics measurements ($VD_t$), respectively. The LSTM network navigates this information utilizing a sequence of gating mechanisms, which are analytically represented in different parts. The \textbf{input gate} controls information flow into the cell state, which can be illustrated as $i_t = \sigma(W_{xi} \cdot x_t + W_{hi} \cdot h_{t-1} + b_i)$. The \textbf{forget gate} as $f_t = \sigma(W_{xf} \cdot x_t + W_{hf} \cdot h_{t-1} + b_f)$ determines which previous information to retain. The \textbf{cell state update} combines new and retained information which can be presented as $g_t = \tanh(W_{xg} \cdot x_t + W_{hg} \cdot h_{t-1} + b_g)$ and $c_t = f_t \cdot c_{t-1} + i_t \cdot g_t$. The \textbf{output gate} produces the final hidden state through $o_t = \sigma(W_{xo} \cdot x_t + W_{ho} \cdot h_{t-1} + b_o)$ and $h_t = o_t \cdot \tanh(c_t)$. In this context, $\sigma$ signifies the sigmoid activation function, with $W$ matrices indicating the acquired weights, $b$ vectors serving as bias components, and $h_{t-1}$ along with $c_{t-1}$ denoting the prior hidden and cell states, respectively.

The PCADS-PI methodology utilizes a trained LSTM network to assess sequential vehicular parameter information and categorizes turning maneuvers into separate types, such as left turns, right turns, and U-turns. The approach to anomaly detection is grounded in probabilistic principles, allowing for the calculation of logarithmic probability scores for each classification of turns based on the observed physical parameters. The process of detection adheres to the following statement.

Given an expected turn type based on system inputs or driver commands, $\log P(\text{Turn}_{\text{Expected}})$ can be calculated and compared against $\log P(\text{Turn}_{\text{Other}})$ for all alternative turn classifications. An anomaly indicator $A'_P$ is initiated upon meeting the condition:
\begin{equation}
\log P(\text{Turn}_{\text{Expected}}) < \log P(\text{Turn}_{\text{Other}})
\end{equation}
The physical behavior of the vehicle appears more aligned with an unexpected type of turn than initially expected, indicating possible system anomalies or the presence of malicious interference. This probabilistic methodology guarantees the detection of anomalies whenever there are substantial deviations in the observed vehicle behavior from the anticipated dynamics, thereby offering a validation mechanism that is grounded in physics. The next section will thoroughly validate this detection method experimentally.
\section{Experiment and Results} \label{sec:experiment}
This section presents the experimental setup and the resulting outcomes to assess the effectiveness of the proposed PCADS model. In order to demonstrate the two stages of the PCADS model, this section is divided into two parts, including the PCADS-CA method and the PCADS-PI method.
\subsection{PCADS-CA Model}
\begin{figure}
    \centering
    \includegraphics[width=1\linewidth]{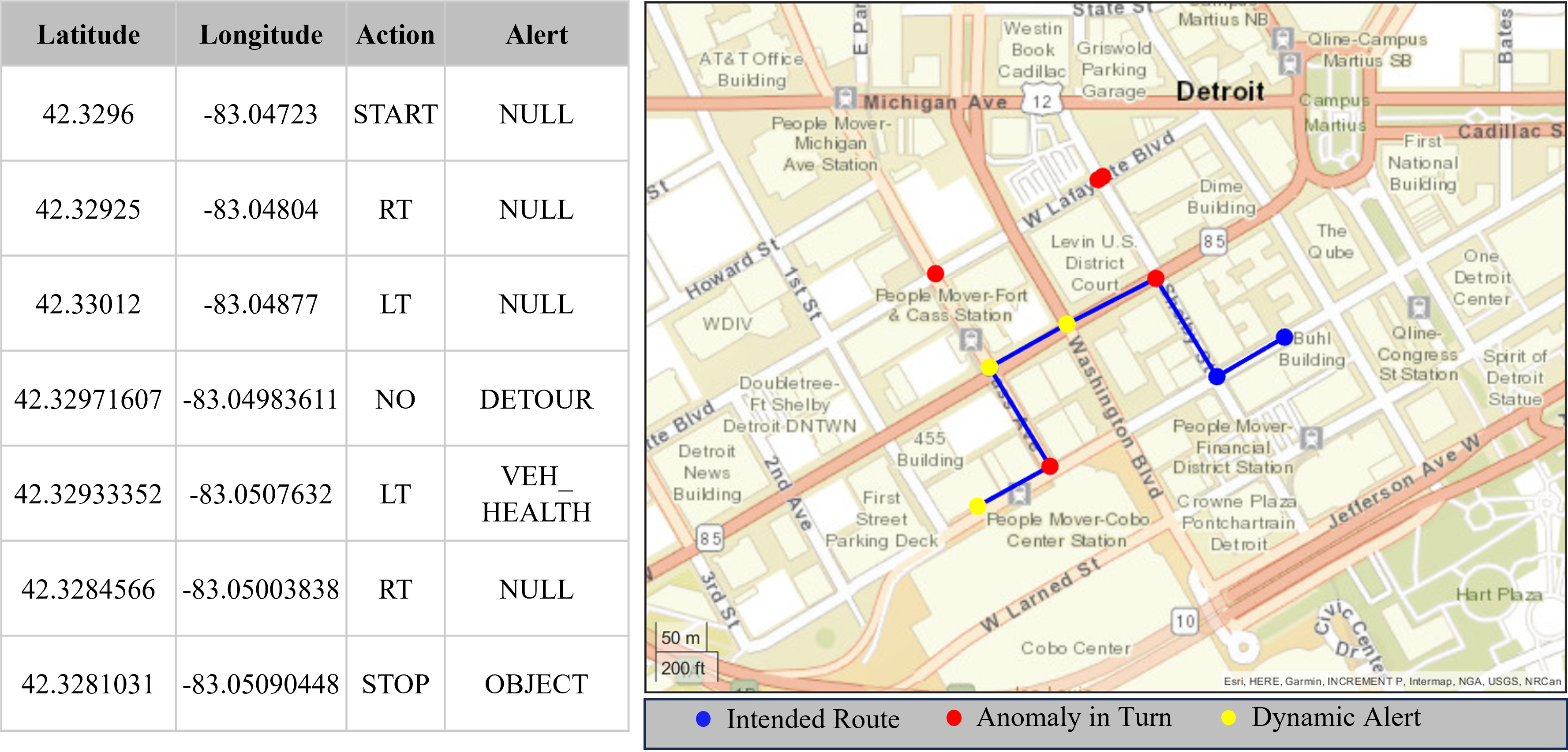}
    \caption{Results for the PCADS-CA incorrect maneuvers.}
    \label{fig:PCADS_CA_Rslt}
\end{figure}

The experiment with the PCADS-CA stage has three steps. First, an original route plan for a delivery vehicle is created with a web-based route planning service from \textit{MyRouteOnline}~\cite{MyRouteOnline}. This provides the IM at each intersection along the delivery route derived from DCs. This corresponds to IM and DC processes in Fig.~\ref{ToD_AD_Framework}. Secondly, the ToD input is created by altering the original route, which includes the expected turn at certain intersections along the route and the modification of the expected time window of the turn. Additionally, an input is added to indicate if there is any dynamic alert on a particular intersection. In the final step, IM from step 1 and altered ToD input and dynamic alert from step 2 are passed to the PCADS-PI algorithm proposed in Section~\ref{ssec:PCADS-CA}. This algorithm is implemented using a MATLAB script such that the results of this experiment are presented in Fig.~\ref{fig:PCADS_CA_Rslt}.
According to this figure, the table shows the DC including the original intended maneuvering action for intersections along the selected delivery route. The most column shows the dynamic alert for an intersection. The diagram shows the results of the PCADS-CA detection. According to this diagram, the blue line indicates the original route and blue dots denote the intersections along the route. The red dots indicate that the PCADS-CA method's detected anomalies in an actual maneuver from the intended action. However, the results also notify about the dynamic alerts at intersections along the route. Generally, the results illustrate that the PCADS-CA model can detect the first stage of anomalies based on the DC. Further, it also provides notification of dynamic alerts to reduce FPs. In the next part, experiments and results for the second stage of the PCADS model are discussed. 
\subsection{PCADS-PI Model}
This section elaborates on the experiment and results with the physics-informed AD stage of the PCADS model. This experiment has two primary steps, including data generation and the AD process, that are described below.
\subsubsection {Data Generation} \label{datagen}
\begin{figure}
    \centering
    \includegraphics[width=1\linewidth]{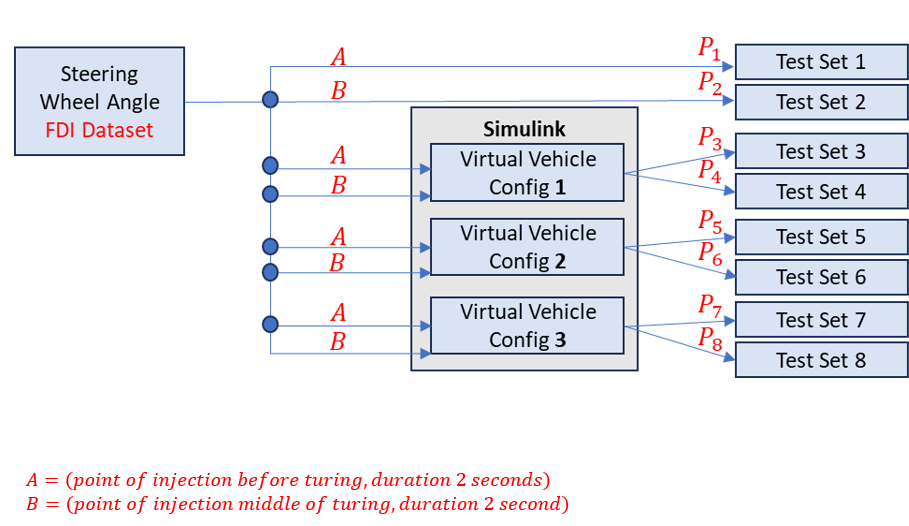}
    \caption{Results for the PCADS-CA incorrect maneuvers.}
    \label{fig:virtual_veh_mdl}
\end{figure}

\begin{figure}[!t]
\centerline{\includegraphics[width=0.9\columnwidth]{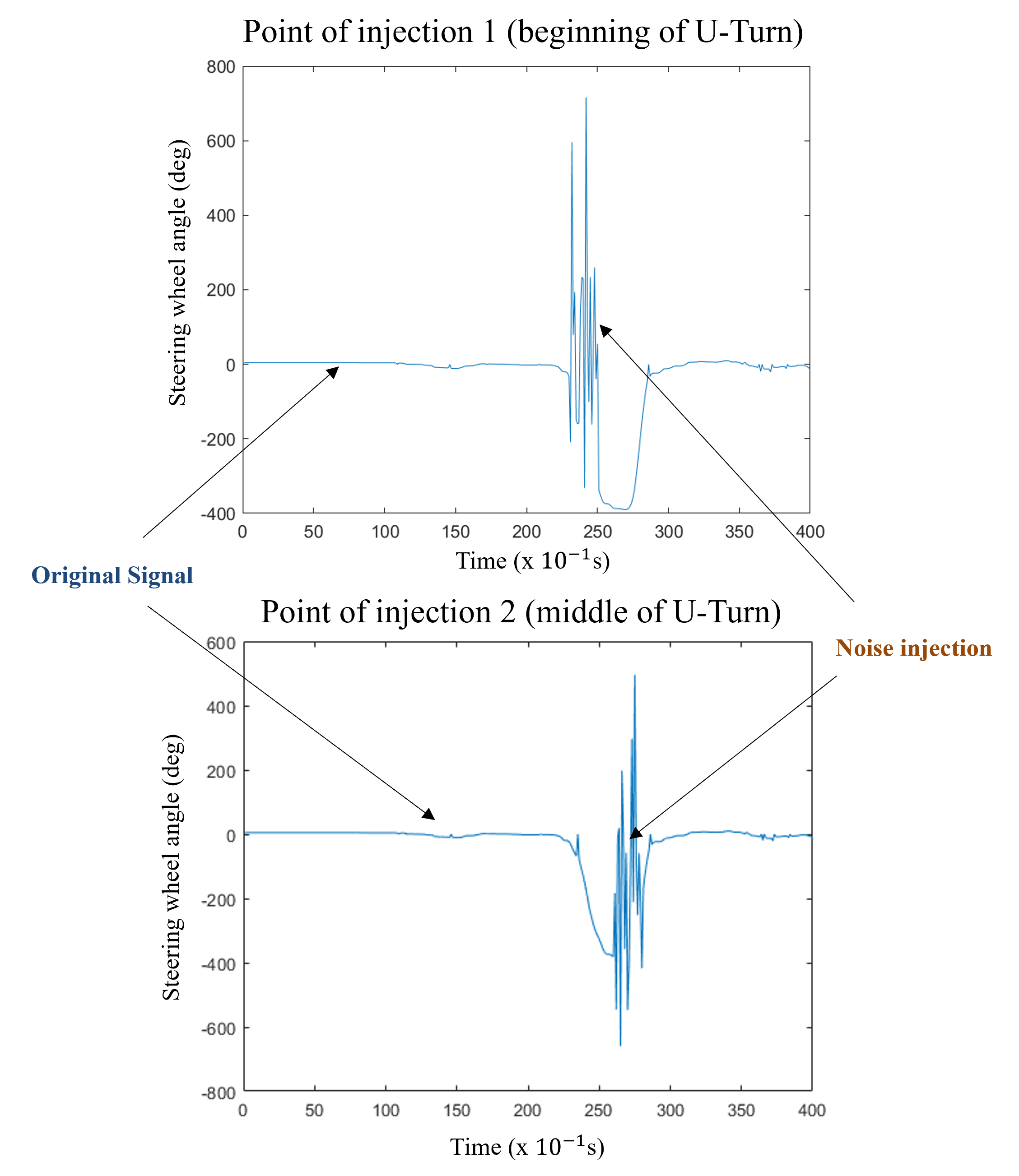}}
\caption{An attack on the steering wheel command.}
\label{fig:AtkStrCmdWoW}
\end{figure}

An experimental dataset for the PCADS-PI model is generated based on the real dataset known as ``D2CAV.'' The dataset contains 75 left turn, 78 right turn, and 62 U-turn scenarios. As per the scope of this paper, steering wheel angle, accelerator pedal, and brake pedal signals are extracted from this dataset. The signals are recorded every 100 ms. This dataset is referred to as a good ToD input dataset. In the next step, this dataset is used as inputs to simulate virtual vehicle models. One dataset of the vehicle physical parameters is created with the good ToD input dataset and another is created for the attack dataset as an input. The data generation of vehicle parameters with steering wheel angle FDI injection is illustrated in Fig.~\ref{fig:virtual_veh_mdl}. In this diagram, steering wheel input is shown as FDI noise at two different points, A and B during the turning action. It can be noted that the good data generation for vehicle physical parameters is a similar process except for the FDI in any input. The configuration of the virtual vehicle model and the set of vehicle physical parameters recorded by simulating the virtual vehicles are the same for good ToD input and ToD input with noise. To generate the good dataset of vehicle physical parameters, the virtual vehicle model is simulated without the good ToD input dataset. The virtual vehicle models are selected from three potential electric drive train configurations with six degrees of freedom. Virtual Vehicle Config 1, 2, 3 refer to a single motor, dual motor, and quad motor used as propulsion motors. Vehicle physical parameters are selected from three subsystems including the energy storage system, traction motor, and vehicle dynamics. The virtual vehicle model is configured and simulated in MATLAB/SIMULINK software. To inject the noise, the attack formula, shown in Eq.~(\ref{eq:Attackformulation}) is implemented using MATLAB. As illustrated in Fig.~\ref{fig:AtkStrCmdWoW}, an attack dataset consists of two points of injection in the steering wheel angle command and the duration of the injected noise is 2 s. The points of injection are at the beginning of turns and during the mid-point of turning. The attack dataset is created by injecting noise into the steering wheel angle in 30 randomly selected observations of the left turn, right turn, and U-turn scenarios. In the final step of data generation, the good dataset and attack dataset of vehicle physical parameters generated from the virtual vehicle model are formatted to train and test the anomaly detection model.
\subsubsection {Anomaly Detection}
An anomaly in trajectory patterns of turning maneuvers is formulated as a sequence to a classification problem. For that reason, the experiment is divided into two steps. Initially, the experiment is conducted to train the ML model with a good dataset and predict 3 classes (i.e., left turn, right turn, and U-turn). A tree-based classifier and 7 NN architectures are trained using the MATLAB DL tool in this case, and the performance is evaluated with standard metrics (i.e., accuracy, precision, recall, F1-score). A value for each metric can range between 0 and 1, where a higher value shows better performance, and the results are presented in Tables~\ref{EXP-Results}. According to this table, the LSTM performance is as follows: minimum accuracy: 0.95, lowest precision: 0.83, lowest recall: 0.89 and lowest F1-score: 0.91. This shows that LSTM predicted left turn, right turn, and U-turn with higher true positive values and true negative values as compared to other neural network architectures employed in this experiment. Based on this observation, an LSTM algorithm is chosen as a base model for the PCADS-PI method.
\begin{table}[!h] 
\caption{\centering A Comparison of ML algorithms for good dataset.}
\label{EXP-Results}
\begin{tabular}{|p{11mm}|c|c|c|c|c|}
\hline
\makecell{Classifier \\ Type}                 & Class      & Accuracy & Precision & Recall & F1-score \\ \hline
\multirow{3}{*}{\makecell{Tree}}           & Left Turn  & 0.85     & 0.78      & 0.83   & 0.80     \\ \cline{2-6} 
                                & Right Turn & 0.94     & 0.92      & 0.92   & 0.92     \\ \cline{2-6} 
                                & U-Turn     & 0.89     & 0.84      & 0.77   & 0.81     \\ \hline
\multirow{3}{*}{\makecell{Narrow \\  NN}}         & Left Turn  & 0.80     & 0.73      & 0.68   & 0.70     \\ \cline{2-6} 
                                & Right Turn & 0.89     & 0.83      & 0.88   & 0.86     \\ \cline{2-6} 
                                & U-Turn     & 0.86     & 0.76      & 0.76   & 0.76     \\ \hline
\multirow{3}{*}{\makecell{Medium \\ NN}}         & Left Turn  & 0.87     & 0.88      & 0.75   & 0.81     \\ \cline{2-6} 
                                & Right Turn & 0.92     & 0.84      & 0.97   & 0.90     \\ \cline{2-6} 
                                & U-Turn     & 0.91     & 0.85      & 0.82   & 0.84     \\ \hline
\multirow{3}{*}{\makecell{Wide \\ NN}}           & Left Turn  & 0.87     & 0.86      & 0.75   & 0.80     \\ \cline{2-6} 
                                & Right Turn & 0.91     & 0.82      & 0.97   & 0.89     \\ \cline{2-6} 
                                & U-Turn     & 0.93     & 0.91      & 0.84   & 0.87     \\ \hline
\multirow{3}{*}{\makecell{Bi-layered \\ NN}}  & Left Turn  & 0.81     & 0.74      & 0.69   & 0.72     \\ \cline{2-6} 
                                & Right Turn & 0.89     & 0.82      & 0.88   & 0.85     \\ \cline{2-6} 
                                & U-Turn     & 0.88     & 0.80      & 0.79   & 0.80     \\ \hline
\multirow{3}{*}{\makecell{Tri-layered \\ NN}} & Left Turn  & 0.82     & 0.75      & 0.72   & 0.73     \\ \cline{2-6} 
                                & Right Turn & 0.87     & 0.83      & 0.82   & 0.83     \\ \cline{2-6} 
                                & U-Turn     & 0.86     & 0.74      & 0.79   & 0.77     \\ \hline
\multirow{3}{*}{\makecell{\textbf{LSTM}}}           & Left Turn  & 0.99     & 0.99      & 0.99   & 0.99     \\ \cline{2-6} 
                                & Right Turn & 0.95     & 0.99      & 0.89   & 0.94     \\ \cline{2-6} 
                                & U-Turn     & 0.95     & 0.83      & 0.99   & 0.91     \\ \hline
\end{tabular}
\end{table}

The resulting data is subjected to an in-depth analysis, the aim of which is to determine the probability density score associated with both correctly and incorrectly predicted observations. An illustrative instance of this analytical approach is exhibited within Table~\ref{classLogp}.

\begin{table}[!h]
\centering
\caption{\centering The LSTM prediction probability score.}\label{classLogp}
\begin{tabular}{|c|c|c|c|c|c|}
\hline
\textbf{Left Turn}                 & \textbf{Right Turn} & \textbf{U-Turn}                 & \textbf{Prediction} & \textbf{Test}  & \textbf{Status} \\ \hline
{ 0.05} & \textcolor{green}{0.92}        & 0.03                        & `RT'          & `RT'  &   \textcolor{green}{TRUE}       \\ \hline
{ \textcolor{green}{0.86}} & 0.08        & 0.07                        & `LT'          & `LT'  &   \textcolor{green}{TRUE}       \\ \hline
{ 0.05} & 0.03        & \textcolor{green}{0.91}                        & `UT'          & `UT'  &   \textcolor{green}{TRUE}       \\ \hline
{ 0.10} & 0.85        & \textcolor{red}{0.04}                        & `RT'          & `UT'  &   \textcolor{red}{FALSE}       \\ \hline
{ 0.05} & 0.93        & \textcolor{red}{0.02}                        & `RT'          & `UT'  &   \textcolor{red}{FALSE}       \\ \hline
{ 0.94} & \textcolor{red}{0.01}        & 0.05                        & `LT'          & `RT'  &   \textcolor{red}{FALSE}       \\ \hline
\end{tabular}
\end{table}
It might be noted that the probability score ranges from 0 to 1. As shown in Table~\ref{classLogp}, the highest probability score observed by an ML classifier for a particular class is reported as a predicted class. When the predicted class matches the true class provided in a test sample, the prediction is TRUE (shown with a green color) while the prediction is FALSE when the probability score of the true class is not the highest one (shown with a red color). This observation verifies that the  LSTM model is able to capture the temporal dependencies and effectively learns the pattern of physical parameters for a valid maneuver. The probability score analysis also shows that the model lowers the probability score when the sequence is anomalous.
In the final step, the trained LSTM model with a good dataset is tested with the attack dataset generated in Section~\ref{datagen} and the results are presented in Table~\ref{tab:ADR}. 
\begin{table*}[!h]
\centering
\caption{\centering An AD rate against 8 test sets (TSs) of FDI attacks.}\label{tab:ADR}
\begin{tabular}{|c|c|c|c|c|c|c|c|c|}
\hline
             TS\# & A.1 & A.2 & B.1 & B.2 & C.1 & C.2 & D.1 & D.2            \\ \hline
\makecell{No. of \\anomaly \\detected} & 10 & 9    & 16       & 15     & 20      & 11      & 30 & 30 \\ \hline  
\makecell{No. of \\FDI tested} & 30 & 30    & 30       & 30     & 30      & 30      & 30 & 30 \\ \hline
ADR (\%) & 33.33 & 30    & \textbf{53.33}       & 50     & \textbf{66.67}      & 36.67      & \textbf{100} & 33.33 \\ \hline
\end{tabular}
\end{table*}
Each TS\# column provides an anomaly detection rate (ADR) for a set of tests with 30 samples of FDI turning maneuvers. `A', `B', `C' and `D' indicate what kind of data is used to train the model in TS\#. `A' means the trained LSTM model with a ToD input. `B' shows the trained LSTM model with vehicle physical parameters for a single motor electric vehicle. The trained LSTM model with vehicle physical parameters for dual and quad motor electric vehicles is mentioned by `C' and `D', respectively. The number 1 in TS\# indicates the point of noise injection at the beginning of turning and 2 means noise is injected at the middle of the turning. As detailed in Table~\ref{tab:ADR}, the Anomaly Detection Rate (ADR) progressively increases with the complexity of the vehicle's drivetrain model. The model trained on single-motor parameters (TS B.1, 53.33\% ADR) outperforms the baseline ToD input model (TS A.1, 33.33\% ADR). This performance is further enhanced with dual-motor parameters (TS C.1, 66.67\% ADR) and culminates in a 100\% ADR for the quad-motor configuration (TS D.1). This demonstrates that parameters related to the drivetrain—specifically individual motor torque, speed, and power consumption—are the most critical for detection. These parameters provide a high-fidelity, difficult-to-spoof fingerprint of the vehicle's physical state. A quad-motor configuration offers the most granular data, as an attacker would need to simultaneously spoof the complex, differential torque distribution across all four motors to remain undetected, which is a significantly more challenging task.

\section{Conclusions and Future Directions} \label{sec:conclusion} 
A foundational framework for the cyber-physical security of ToD systems was established by this paper. Through a TARA, FDI attacks on steering commands were identified as a high-risk vulnerability. Subsequently, a novel attack model was developed, and, most critically, a PCADS was proposed and validated to mitigate this threat. The core hypothesis of this work is validated by the experimental results. A model based solely on control inputs is outperformed by a physics-informed detection model that uses an LSTM architecture, which is adept at capturing temporal patterns, to effectively learn the physical signature of a valid maneuver from vehicle physical parameters.

A foundation for research to develop robust and intrinsic security layers for ToD, extending the Defense-in-Depth (DiD) paradigm into the vehicle's physical domain, is provided by the contributions presented here—the ToD threat model, the FDI attack formulation, and the PCADS framework. This is critical for a comprehensive cyber-physical security roadmap. As part of future work, a holistic dataset for FDI attacks on ToD control with various combinations of noise, points of injection, and attack durations is planned to be developed. As an extension of the PCADS model, other AD approaches will be explored, and a comparative performance analysis will be carried out on this extensive attack dataset.
\section{Appendix} \label{appendix-1}
\subsection{Data Availability and Supplementary Material}
\label{appendix}

The source codes, datasets, and additional supplementary materials supporting the findings of this study are publicly available through the GitHub repository (\href{https://github.com/ghostsubha/TODS_LMD_AD}{https://github.com/ghostsubha/TODS\_LMD\_AD}). This repository contains comprehensive implementation details, experimental scripts, datasets, and detailed instructions for reproducing all results.
\bibliographystyle{IEEEtran}
\bibliography{IEEEabrv,RefDatabase}

\begin{IEEEbiography}[{\includegraphics[width=1in,height=1.25in,clip,keepaspectratio]{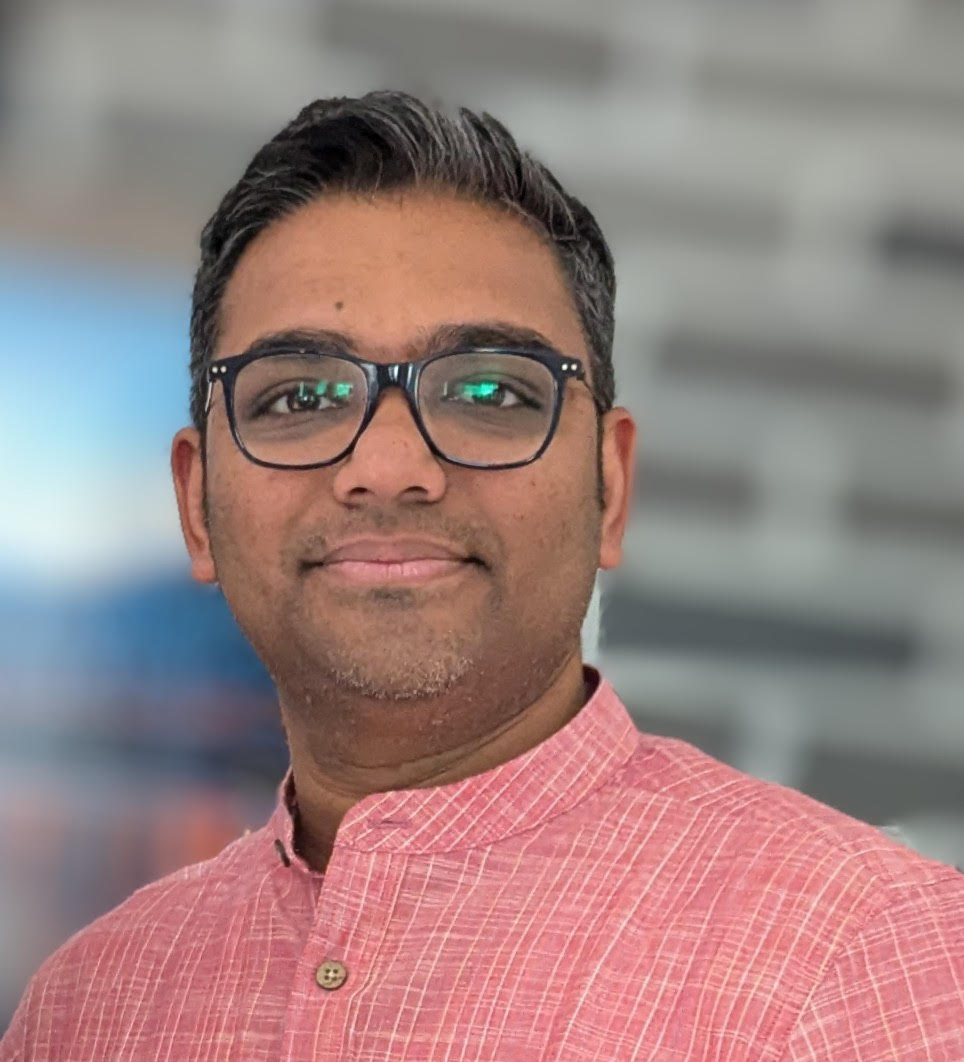}}]{Subhadip Ghosh} (M'21--SM'24) He is an automotive system-software engineer at Ford Motor Company, MI, USA. He received a Bachelor of Technology in Computer Science and Engineering from West Bengal University of Technology, India, in 2006 and an MS in Electric-Drive Vehicle Engineering from Wayne State University, Michigan, in 2014. Since 2006, he has worked in systems and software development for automotive ECUs in the body, EV, and ADAS domains. He has provided technical leadership in core product development, foundational architecture, and advanced feature design. In 2024, he has received a Doctor of Engineering degree in Automotive Systems and Mobility from the University of Michigan-Dearborn, USA. His research interest is in the cyber-security and cyber-physical security of automated and connected vehicles.
\end{IEEEbiography}

\begin{IEEEbiography}[{\includegraphics[width=1.0in,height=1.5in,clip,keepaspectratio]{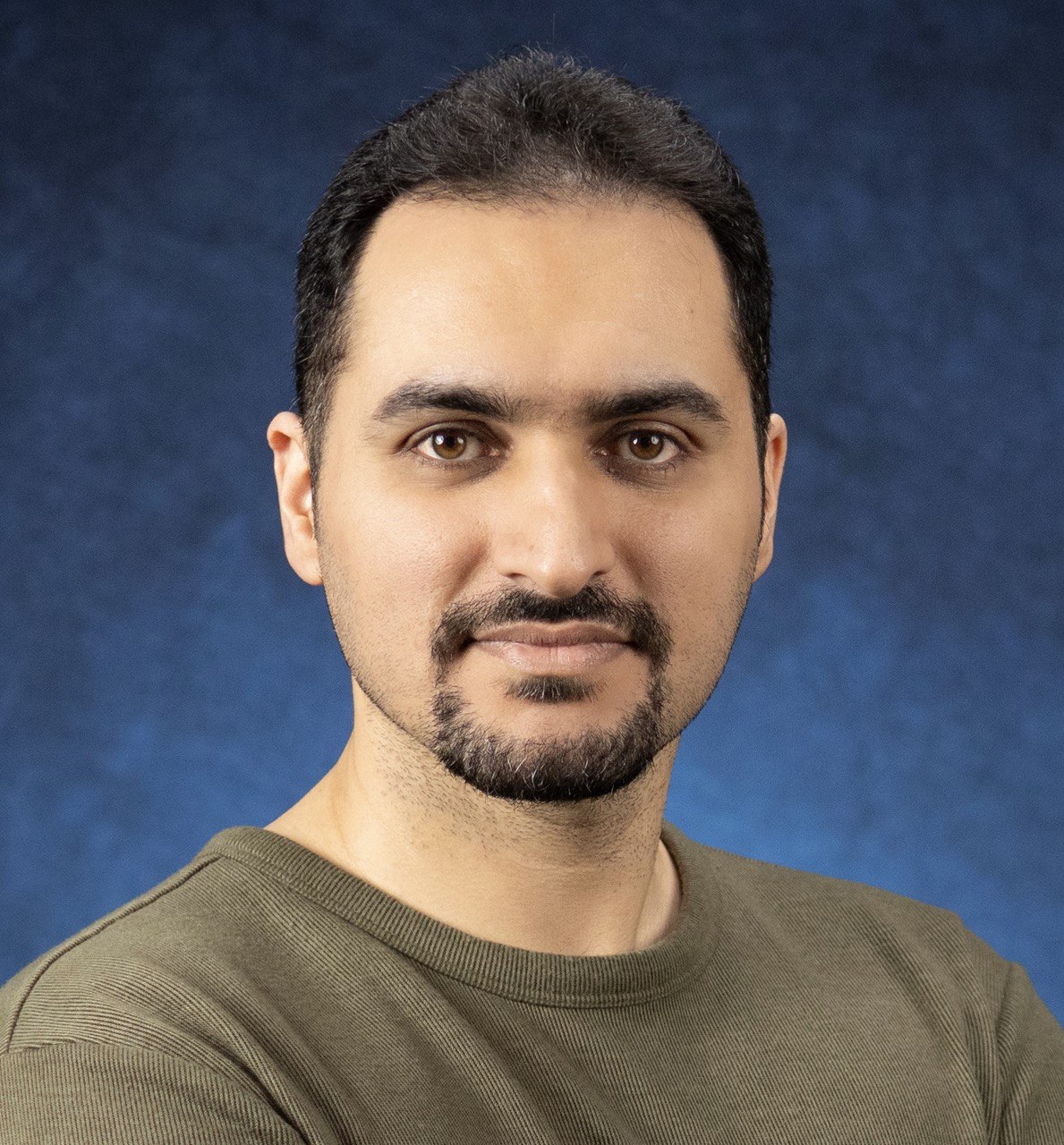}}]
{Aydin Zaboli} (GSM'21) is currently pursuing a Ph.D. degree in electrical, electronics, and computer engineering at the University of Michigan–Dearborn, Dearborn, MI, USA. His research focuses on smart grid security, autonomous vehicles, anomaly detection, transportation electrification, renewable energy resources, and load forecasting. He has served as a reviewer for more than 250 papers in prestigious journals and conferences, particularly IEEE Access, IEEE Transactions on Transportation Electrification, IEEE Transactions on Vehicular Technology, and IEEE Transactions on Smart Grid, contributing to the advancement of research in smart grids and transportation electrification. He is also the recipient of the Rackham Predoctoral Fellowship from the University of Michigan–Rackham Graduate School for the academic year 2024–2025.
\end{IEEEbiography}

\begin{IEEEbiography}[{\includegraphics[width=1in,height=1.25in,clip,keepaspectratio]{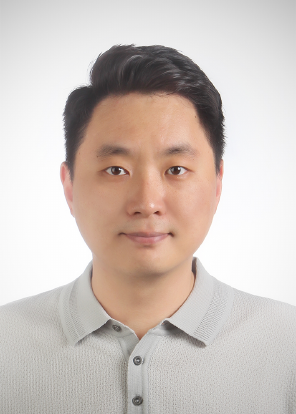}}]{Junho Hong} (M'14--SM'22) He is an associate professor in the Department of Electrical and Computer Engineering at the University of Michigan–Dearborn. He received his Ph.D. degree with Cyber-security of Substation Automation System in Electrical Engineering from Washington State University, Pullman, in 2014. During 2014–2019, he worked with ABB where he provided technical project leadership and supported strategic corporate technology development/productization in areas related to cyber-physical security for substations, power grid control and protection, renewable integration, and utility communications. He has been working on the cyber-security of energy delivery systems with the Department of Energy (DOE) as Principal Investigator (PI) and Co-PI in the areas of substations, microgrids, HVDC, FACTS, and high-power EV chargers. He serves in Cigre WG D2.50, ``Electric power utilities’ cyber-security for contingency operations.''
\end{IEEEbiography}

\begin{IEEEbiography}[{\includegraphics[width=1in,height=1.25in,clip,keepaspectratio]{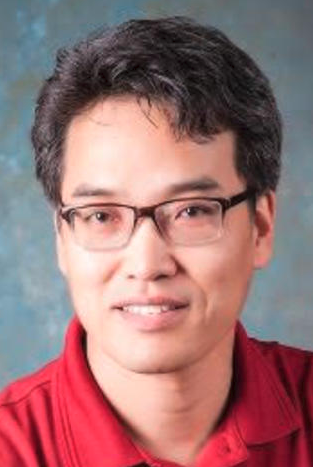}}]{Jaerock Kwon}(M'06--SM'20) received his B.S. and M.S. degrees from Hanyang University, Seoul, Korea, in 1992 and 1994, respectively. Between 1994 and 2004, he worked for LG Electronics, SK Teletech, and Qualcomm Internet Services. Then, he received his Ph.D. degree in computer engineering from Texas A\&M University, College Station, USA, in 2009. From 2009 to 2010, he was a professor in the Department of Electrical and Computer Engineering at Kettering University, Flint, MI, USA. Since 2010, he has been a professor in the Department of Electrical and Computer Engineering at the University of Michigan—Dearborn, MI, USA. His research interests include mobile robotics, autonomous vehicles, and artificial intelligence. Dr. Kwon’s awards and honors include the Outstanding Researcher Award, Faculty Research Fellowship (Kettering University), and SK Excellent Employee (SK Teletech). He served as President of the Korean Computer Scientists and Engineers Association in America (KOCSEA) in 2020 and 2021.
\end{IEEEbiography}
\EOD
\end{document}